\documentclass[aps, prd, reprint, showpacs, superscriptaddress, twocolumn, nofootinbib]{revtex4-1}    % for review and submission
\usepackage{graphicx}	% Include figure files
\usepackage{graphics}
\usepackage{dcolumn}	% Align table columns on decimal point
\usepackage{bm}		% bold math
\usepackage{tabularx} 	% extra features for tabular environment
\usepackage{amsmath}    % improve math presentation
\usepackage{amssymb}
\usepackage{verbatim}
\usepackage{hyperref}
\usepackage{cleveref}
\usepackage{lipsum}
\usepackage{braket}
\usepackage{physics}
\usepackage{color}
\usepackage{float}
\usepackage{natbib}
\usepackage{color}
\usepackage{xcolor}
\begin{document}

\title{Neutrino Burst-Generated Gravitational Radiation From Collapsing Supermassive Stars}

\author{Jung-Tsung Li}
\email{jul171@ucsd.edu}
\affiliation{Department of Physics, University of California, San Diego, California, 92093, USA}

\author{George M. Fuller}
\email{gfuller@ucsd.edu}
\affiliation{Department of Physics, University of California, San Diego, California, 92093, USA}

\author{Chad T. Kishimoto}
\email{ckishimoto@sandiego.edu}
\affiliation{Department of Physics, University of California, San Diego, California, 92093, USA}
\affiliation{Department of Physics and Biophysics, University of San Diego, California, 92110, USA}

%------------------------------------------------------------------------------------------------------------------------
% Abstract
%------------------------------------------------------------------------------------------------------------------------

\begin{abstract}
We estimate the gravitational radiation signature of the electron/positron annihilation-driven neutrino burst accompanying the asymmetric collapse of an initially hydrostatic, radiation-dominated supermassive object suffering the Feynman-Chandrasekhar instability. An object with a mass $5\times10^4\,M_\odot<M<5\times10^5\,M_\odot$, with primordial metallicity, is an optimal case with respect to the fraction of its rest mass emitted in neutrinos as it collapses to a black hole: lower initial mass objects will be subject to scattering-induced neutrino trapping and consequently lower efficiency in this mode of gravitational radiation generation; while higher masses will not get hot enough to radiate significant neutrino energy before  producing a black hole. The optimal case collapse will radiate several percent of the star's rest mass in neutrinos and, with an assumed small asymmetry in temperature at peak neutrino production, produces a characteristic linear memory gravitational wave burst signature. The timescale for this signature, depending on redshift, is $\sim1{\rm~s}$ to $10{\rm~s}$, optimal for proposed gravitational wave observatories like DECIGO. Using the response of that detector, and requiring a signal-to-noise ratio SNR $>$ 5, we estimate that collapse of a $\sim 5\times10^4\,M_\odot$ supermassive star could produce a neutrino burst-generated gravitational radiation signature detectable to redshift $z\lesssim7$. With the envisioned ultimate DECIGO design sensitivity, we estimate that the linear memory signal from these events could be detectable with SNR $> 5$ to $z \lesssim13$.
\end{abstract}


\maketitle

%------------------------------------------------------------------------------------------------------------------------
% Section 1, Introduction
%------------------------------------------------------------------------------------------------------------------------
\section{Introduction}\label{introduction}

In this paper we point out a surprising and serendipitous connection between the weak interaction physics of supermassive star (SMS) collapse to a black hole, the generation of a linear memory gravitational radiation signature from the neutrino burst that ensues in such an event, and the detection of this signature in proposed mid-frequency ($\sim 0.1\,{\rm Hz}$) gravitational wave observatories like DECIGO~\cite{Seto:2001qf, Kawamura:2011zz}.

We consider stars with a range of masses which falls into the category of the classic SMS of Fowler and Hoyle \cite{Foyle:1963aa, Iben:1963, Fowler:1964zza, Zeldovich:1969sb, Fuller:1986}, {\it i.e.}, $M\gtrsim10^4\,{M_\odot}$. These are initially hydrostatic, high entropy, fully convective configurations, with the bulk of the mass-energy provided by baryon rest mass, but with the entropy and pressure support stemming predominately from the radiation field. The result is a star with adiabatic index perilously close to $4/3$, trembling on the verge of instability, and therefore ripe for destabilization by tiny (in this case) nonlinear effects inherent in General Relativity: the Feynman-Chandrasekhar post-Newtonian instability~\cite{Chandrasekhar:1964zz, Feynman:1996kb}. There are many unresolved issues surrounding the formation and existence of such objects, their fate once they suffer the post-Newtonian instability, and the weak interaction and nuclear reaction history of the material in the collapsing star and the associated neutrino emission. We will not address these issues here, except insofar as they impact our key requirement for this work: an ultra-high entropy star that collapses to a black hole but remains essentially transparent to neutrinos until nearly the bitter end, when a black hole forms.

First, it must be pointed out that there exists no direct observational evidence, or even an indirect nucleosynthesis or chemical evolution argument that these stars ever existed. Moreover, even the question of whether nature could produce such objects remains unanswered. The existence of supermassive black holes (SMBHs) powering quasars at high redshift is indisputable~\cite{Mortlock:2011va, Venemans:2013npa, Wu:2015, Banados:2017, 2017arXiv171201886V}. This has long invited speculation on the origin of these SMBHs and about the masses of \lq\lq seed\rq\rq\ black holes from which early SMBH could arise via mergers.

Begelman \& Rees~\cite{Begelman:1978, Rees:1984si} drafted a flow chart showing the routes to SMBH formation. Several of these include the formation of a SMS, either by suppression of fragmentation of a collapsing primordial gas cloud~\cite{Haehnelt:1993yy, Loeb:1994wv, Eisenstein:1994nh, Oh:2001ex, Bromm:2002hb, Koushiappas:2003zn, Lodato:2006hw}, or by forming such an object in a dense star cluster via tidal gas stripping in close stellar encounters or by collisions~\cite{shapiro1985collapse, quinlan1987collapse, quinlan1989dynamical, PortegiesZwart:2004ggg}. In the primordial gas cloud collapse scenario, the outcome may depend on the gas accretion rate. High accretion can lead to a non-adiabatic configuration, essentially a compact object at the center with a distended lower density envelope~\cite{Begelman:2006db, Begelman:2010ac}. This will not produce the high entropy, fully convective (adiabatic) configuration we consider here. It is an open question whether a lower gas accretion rate, plus fragmentation suppression through heating or reduced cooling, can lead to this result. Certainly, stellar disruption or collision could produce a high entropy self-gravitating star, but this object might have relatively higher metallicity and therefore may explode via hot CNO hydrogen burning rather than collapse to a black hole~\cite{Fuller:1986}.

Instead, we focus on the classic primordial metallicity hydrostatic SMS, completely convective, where the density, temperature, and pressure runs with radius are well described by a Newtonian index $n=3$ polytrope, at least initially, prior to the onset of the post-Newtonian instability. Moreover, we point out here that if such a high entropy hydrostatic SMS {\it did} form, its subsequent collapse and neutrino emission can produce a unique gravitational wave burst signature, potentially detectable even for a collapse at very high redshift.

Gravitational radiation originating in the collapse of a rapidly rotating SMS to a black hole has been studied in Ref.~\cite{Shibata:2016vzw}. That study found that  most of the energy radiated in gravitational waves in SMS collapse is generated either by the time changing quadrupole moment of the baryons before trapped surface formation, or subsequent black hole ring down, all depending on the initial angular momentum content of the star.
The gravitational wave signal produced this way will be a conventional, oscillatory one, well matched to proposed detectors in the mid- to low-frequency band.

Here we consider something quite different, both in the origin of the gravitational radiation in a SMS collapse and in the nature and signature of this radiation in detectors. 
While the study in Ref.~\cite{Shibata:2016vzw} centered on the effects of the bulk of the mass-energy, the baryons, in these objects, here we examine a complementary issue, the role of the very sub-dominant neutrino component in gravitational radiation production. Gravitational waves generated by the neutrino burst associated with ordinary core collapse supernovae and neutron star production is an old and well investigated idea~\cite{Burrows:1995bb, Mueller:1997, Fryer:2004wi, Ott:2006qp, Dessart:2006gd, Kotake:2006aq}, but these venues are eventually opaque to neutrinos and involve neutrino emission from a neutrino sphere. By contrast, we examine what happens in a star with significant neutrino emission, more than a few percent of the gravitational binding energy, yet has high enough entropy and therefore low enough density to be essentially transparent to neutrinos until near gravitational trapped surface formation. We will show how, unlike a {\it static} neutrino-transparent configuration, a collapsing but otherwise transparent SMS can \lq\lq lock in\rq\rq\ an asymmetry in neutrino emissivity and thereby generate a neutrino burst with a time-changing effective quadrupole moment.

Interestingly, since neutrinos emitted during the collapse are gravitationally unbound, the accompanying gravitational radiation generated by the neutrino burst will constitute what is termed gravitational waves with linear memory (GWM)~\cite{zel1974radiation, Smarr:1977fy, Kovacs:1978eu, Braginskii:1985}. A GWM is a non-oscillatory gravitational wave that leaves a net change in the gravitational wave strain after the signal has passed by.

The GWM effect was first discussed in its linear form in the 1970-80s~\cite{zel1974radiation, Smarr:1977fy, Turner:1977, Epstein:1978dv, Turner:1978jj, Kovacs:1978eu, Braginskii:1985, Braginskii:1987}. For a recent review see Ref.~\cite{Favata:2010zu}. In general, systems with more than one mass component gravitationally unbound to each other can produce gravitational waves with linear memory. Several such production mechanisms have been discussed, for example, hyperbolic binaries~\cite{Turner:1977, Kovacs:1978eu}, gamma-ray bursts~\cite{Sago:2004pn}, matter ejecta from supernova explosions~\cite{Burrows:1995bb, Mueller:1997, Fryer:2004wi, Ott:2006qp, Dessart:2006gd, Kotake:2006aq}, and anisotropic neutrino emission~\cite{Turner:1978jj, Epstein:1978dv}. The prospects for detecting the memory effect have been studied in Refs.~\cite{Braginskii:1985, Braginskii:1987, 300yearsofgravitation}.

This paper is organized as follows.
In Section~\ref{Characteristics_SMS} we discuss the neutrino emission from SMSs.
In Section~\ref{quadrupole_approximation} we estimate the strain magnitude of the gravitational wave signals from this mechanism in collapsing SMSs and calculate the corresponding waveforms.
In Section~\ref{signal to noise ratio} we discuss the prospects for detection of these signals with the next generation space-based gravitational wave detectors.
In Section~\ref{other_sources} we discuss other possible sources which could also produce the linear GWM signal with strain magnitude and timescale similar to those originated from high-redshift SMS collapse.
Throughout this paper we adopt geometric and natural units, $G=c=k_{\rm b}=\hbar=1$, and assume $\Lambda$CDM cosmology with the closure fraction of the non-relativistic component chosen as $\Omega_\text{M} = 0.3$, the vacuum energy contribution to this fraction taken as $\Omega_\Lambda=0.7$, and the Hubble parameter at the current epoch in units of $100\,{\rm km}\,{\rm s}^{-1}\,{\rm Mpc}^{-1}$ taken to be $h=0.7$.

%------------------------------------------------------------------------------------------------------------------------
%  Section 2, Characteristics of SMSs
%------------------------------------------------------------------------------------------------------------------------
\section{Characteristics of SMSs}\label{Characteristics_SMS}

%%--------------------------------
%%     Section 2A
%%--------------------------------

\subsection{Total neutrino energy from the collapsing SMS}\label{neutrino energy}
A hydrostatic, fully convective SMS with stellar mass $M_\text{SMS} \gtrsim5\times10^4\,M_\odot$ has a structure well represented by an $n=3$ polytrope. 
It is radiation dominated and most of its entropy is carried by photons and electron/positron pairs. 
The entropy per baryon in units of Boltzmann's constant $k_{\rm b}$ is typically $s\approx\left({M}/{M_\odot}\right)^{1/2} \approx 300\left({M^\text{HC}_5}\right)^{1/2}$, where $M_5^\text{HC}$ is the homologous core mass in units of $10^5\,M_\odot$~\cite{Shi:1998nd}. As the SMS gradually radiates away energy and shrinks in radius, the star eventually suffers post-Newtonian instability and begins to collapse. For $M^{\rm HC}\sim10^5\,M_\odot$, instability sets in roughly at the onset of hydrogen burning. A fraction, likely a few tens of percent, of the initial stellar mass may collapse homologously, depending on the history of neutrino energy loss, nuclear burning and initial angular momentum content and distribution~\cite{Fuller:1986}. It's this homologous core that produces the initial BH.

The gravitational binding energy liberated in the collapse is $E_s \sim M^{\rm HC}$. Most of this energy is trapped in the BH, but a small fraction will be radiated as neutrinos. These neutrinos are produced mostly via electron/positron pair annihilation into neutrino pairs. Since the rate of the energy emissivity of this neutrino pair production channel is proportional to a high power of the temperature, $\propto T^9$~\cite{Schinder:1986nh, Itoh1989, Itoh1996}, the bulk of the radiated neutrino energy will be produced very close to the BH formation point, where the temperature is the highest. Just how high the plasma temperature gets before redshift associated with gravitational trapped surface formation cuts off neutrino escape depends on details of SMS evolution, e.g., nuclear burning and convective timescales during the collapse, and on the collapse rate near BH formation.

Shi \& Fuller~\cite{Shi:1998nd} estimated that the total neutrino energy emitted during the collapse of a non-rotating SMS as $3.6\times 10^{57}\left(M_5^{\rm HC}\right)^{-0.5}~{\rm ergs}$ in a timescale $\Delta \tau = M_5^\text{HC}~{\rm s}$. More sophisticated hydrodynamic simulations conducted by Linke et al.~\cite{Linke:2001mq} show that the innermost 25\% of the SMS mass will collapse homologously to a BH, emitting neutrinos on a timescale approximately 11 times longer than estimated by Shi \& Fuller. In the mass range $10^5\,M_\odot \lesssim M^{\rm HC} \lesssim 5 \times 10^5\,M_\odot$, they calculate the total energy emitted in neutrinos to be approximately $3\%$ of what was found in Shi \& Fuller. Though still a substantial amount of energy, this result shows a considerable discrepancy with Shi \& Fuller. The difference between these calculations reflect the different ways in which neutrino emission and redshift near the BH formation point were calculated. In turn, this physics is dependent on the treatment of in-fall and collapse timescales, pressure, and the adiabat of collapse. The Linke et al. calculation likely is more realistic, as it gives a self consistent calculation of neutrino emission coupled to collapse dynamics and redshift. Nevertheless large uncertainties remain in the physics \lq\lq upstream\rq\rq\ of the BH formation point. Consequently, we will consider both calculations in our assessment of the neutrino burst-generated linear memory gravitational wave signal from SMS collapse.

The calculations in Ref.~\cite{Shi:1998nd} do not apply for $M^{\rm HC}_5 \lesssim 0.1$, {\it i.e.,} where neutrinos may be trapped via scattering on electrons and positrons. For example, a  homologous core with mass $M^{\rm HC}_5 = 0.1$, close to BH formation, has a neutrino mean free path smaller than the Schwarzschild radius. We conclude that for $M^{\rm HC}_5 \lesssim 0.1$ the homologous core will be subject to neutrino scattering-induced trapping and is opaque to neutrinos. It is likely that a significant fraction of neutrinos will be carried into, and trapped inside, the BH in this case.

At higher SMS masses, a smaller fraction of the SMS rest mass is radiated as neutrinos even though the total gravitational binding energy released in the collapse is higher. This stems from the fact that electron/positron pair annihilation neutrino emissivity scales as the ninth power of the temperature, whereas the temperature scales as $({M^{\rm HC}})^{-1/2}$. At large SMS mass the core will not get hot enough to radiate a significant fraction of $E_s$ before BH formation. When $M^{\rm HC}_5\gtrsim 10$, less than $0.1$ percent of the homologous core gravitational binding energy would be emitted as neutrinos.

Subject to scattering-induced neutrino trapping at low SMS masses, and low neutrino emissivity at high SMS masses, the optimal mass range of the homologous core for a maximal fraction of the SMS mass to be radiated in neutrinos is $5\times10^4\,M_\odot<M^{\rm HC}<5\times10^5\,M_\odot$.

As discussed above, there remain open questions in the evolution and collapse physics of SMSs. These issues can be relevant for the neutrino burst accompanying SMS collapse. In part, uncertainties in the characteristics of the eventual neutrino burst arise from the fact that the total energy, internal plus gravitational, of these objects near their instability points will be very close to zero. Relatively small changes in nuclear burning history or neutrino emission history may lead to significant subsequent alterations in the thermodynamic history of collapse. During the collapse, neutrino emission and escape remove entropy from the star, while nuclear burning in effect counters this by adding entropy. By far the biggest effect is the former, entropy loss, but the latter entropy source helps determine the entropy content relevant for peak neutrino emission just before the BH formation point. The small effect from nuclear burning in making the entropy, and hence temperature, slightly higher can be significant because the neutrino emission rate from electron/positron pair annihilation scales as the ninth power of the temperature. Though there is negative feedback between the competing processes of neutrino engendered entropy loss and added entropy from nuclear burning, in the end nuclear burning will mean stronger neutrino emission overall and a larger fraction of the homologous core rest mass radiated as escaping neutrinos.

That is, as long as the energy production from nuclear burning is not large enough, or not optimally phased in time or location, so as to cause the thermonuclear explosion and disruption of the star! An explosion caused by nuclear burning early in the collapse obviously precludes production of a BH. The lower end of our considered range of SMS masses may be the most vulnerable to the uncertain details of the phasing and interplay of nuclear burning, convection, rotation, and neutrino emission. For example, the calculation reported in Ref.~\cite{Chen:2014yea} suggested that for a narrow range of SMS masses around $5.5\times{10}^4\,M_\odot$ \lq\lq explosive\rq\rq\ helium burning immediately subsequent to the post-Newtonian instability point would be sufficient to cause an explosion of the star, even with primordial metallicity at SMS formation. This calculation highlights the outstanding uncertainties associated with SMS evolution up to the instability point and subsequently.

%%--------------------------------
%%     Section 2B
%%--------------------------------
\subsection{Creating an anisotropic neutrino energy flux}\label{Anisotropic nu emission}

There can be another consequence of electron/positron annihilation-generated neutrino energy emission being proportional to nine powers of temperature~\cite{Schinder:1986nh, Itoh1989, Itoh1996}. Because of this high sensitivity to the temperature, even a small anisotropy in the temperature can translate into an order of magnitude larger neutrino emissivity anisotropy. For example, a configuration with a $2.5\%$ lower temperature at the equator than at the poles will have an approximately $25\%$ neutrino emissivity asymmetry between volume elements along the equatorial plane and the polar direction.

In the SMSs we consider here, the bulk of the pressure, $P$, comes from relativistic particles, implying that $P \propto T^4$. Therefore $\delta P / P = 4 \delta T / T$. So a $2.5\%$ decrease in temperature corresponds to a $10\%$ decrease in pressure. A rotation-driven centrifugal acceleration decreases the required pressure support in the star's equatorial plane relative to its polar direction. Interestingly, $10\%$ difference in pressure between the equator and pole on a 2-sphere near the maximum neutrino emissivity point, in turn, likely would not significantly change the free-fall collapse time there. For $M^{\rm HC} = 10^5\,M_\odot$, an angular speed of $\omega \sim 0.22~{\rm rad}\,{\rm s}^{-1}$ at BH formation, corresponding to dimensionless angular momentum of $J/M^2 \sim 0.18$, would produce a $25\%$ neutrino energy emissivity anisotropy.

If the SMS is both transparent to neutrinos and static, this emissivity asymmetry would not be imprinted on the neutrinos escaping to infinity. The reason for this is simple: Neutrino emission from each volume element in the core will produce a symmetric neutrino emission pattern, radiating neutrinos isotropically, and in a completely neutrino-transparent, spherically symmetric star where each volume element is at rest, the neutrino radiation seen by distant observers will be spherically symmetric and static. However, many of these conditions are violated in a real, collapsing star. A collapsing star, where fluid elements {\it move}, will lock-in some of the temperature variation-created local emission anisotropy discussed above. The mechanism for this is rooted in the non-equivalence of neutrino directions in the collapsing star and, in particular, a direction-dependent differential blueshift-redshift akin to the integrated Sachs-Wolfe (ISW) effect in cosmology~\cite{Sachs:1967er}.

\begin{figure}[t]
\includegraphics[width=0.34\textwidth]{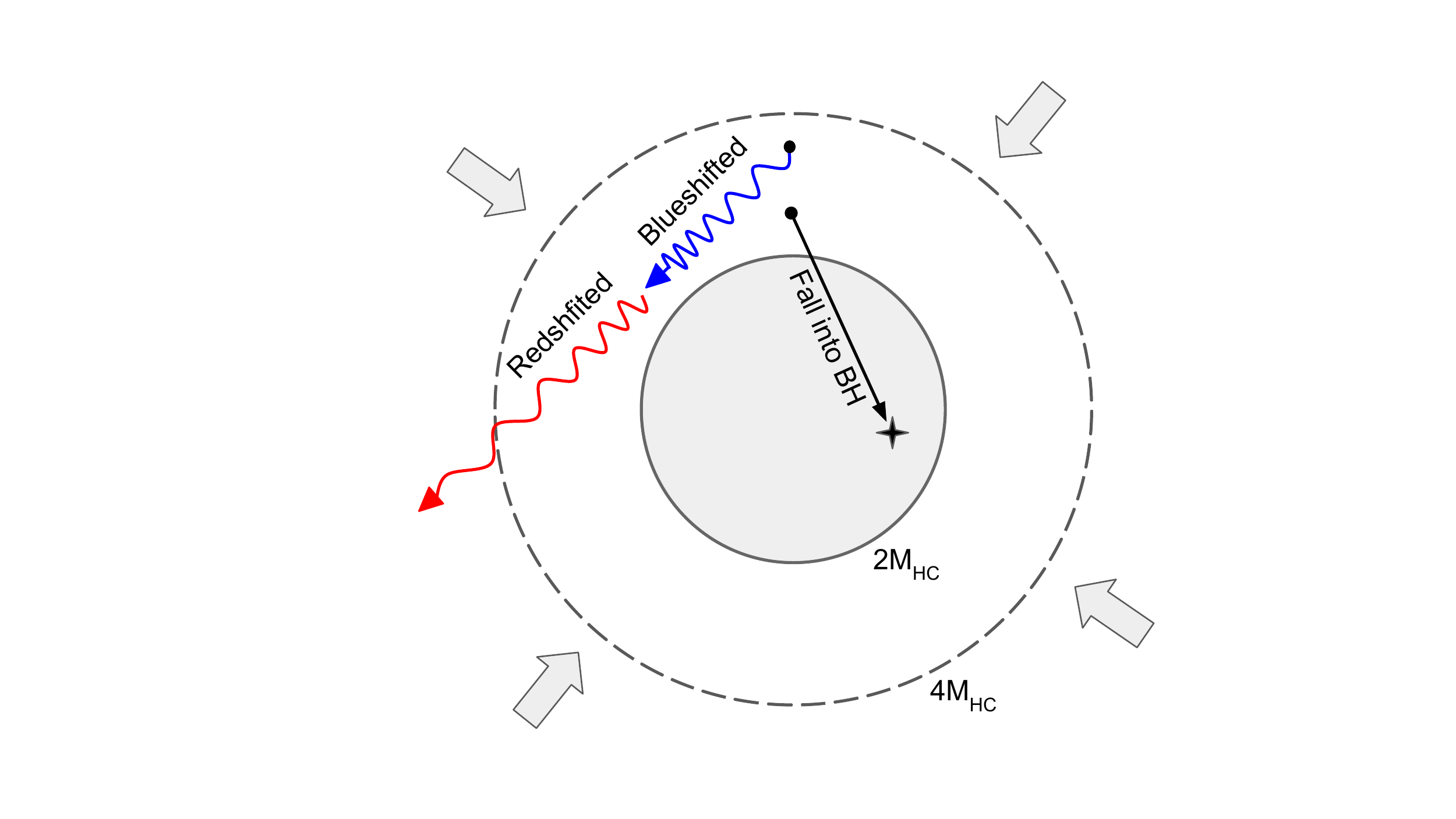}
\caption{Illustration of anisotropic neutrino emission production by the integrated Sachs-Wolfe-like effect. Neutrinos moving toward the central core may fall into the BH or suffer an ISW-like differential blueshift-redshift effect driven by the increase of gravitational potential with time in a collapsing SMS.}
\label{fig:ISW_BH}
\end{figure}

Neutrinos emitted into an inwardly-directed pencil of directions will escape from the star with significantly less energy flux than they were born with. Of course, that is true for any redshifted neutrino, but the point here is that the extent of the unbalanced blueshift-redshift is emission angle-dependent. This breaks spherical symmetry in the neutrino-transparent star. Neutrinos will gain energy, i.e., experience blueshift, as they stream toward the center of the star (homologous core or BH) and lose energy, suffer redshift, as they stream away from the center. However, the key point is that this geometry is not {\it static}, and the SMS is collapsing. In the time frame over which most of the neutrinos are produced, the SMS has significantly collapsed, causing the gravitational potential well to become correspondingly deeper. Consequently, the redshift will be larger than the blueshift. This represents a net gravitational redshift, along with absorption by the BH (the ultimate gravitational redshift) for some neutrino directions, implying that although the emission from a given incremental volume element is isotropic, the redshift and absorption of neutrinos produced by this volume element is not. Fig.~\ref{fig:ISW_BH} illustrates the geometry of this differential blueshift-redshift effect. This is how a net anisotropy in the neutrino emission from the SMS, as observed by a distant observer, can be generated.

In Appendix~\ref{appendix c}, we present an order of magnitude estimate of the transformation of a neutrino emissivity anisotropy, $\eta$, (caused by, {\it e.g.}, modest SMS rotation) to a neutrino energy flux asymmetry, $\alpha$, as measured by a distant observer, by this ISW-like mechanism. We stress that the model in the Newtonian treatment in Appendix~\ref{appendix c} is meant to be an order of magnitude estimate and is not meant to be a detailed analysis of the many general relativistic effects that may affect the result, which is beyond the scope of this work. Nevertheless, our intriguing results suggest that such a fully general relativistic study is warranted.

There are two important results from the Newtonian model in Appendix~\ref{appendix c}: that the neutrino energy flux asymmetry has the opposite sign from the neutrino emissivity asymmetry; and that is an order of magnitude smaller. If a rotation-created temperature asymmetry resulted in an $\eta = 0.25$ neutrino emissivity asymmetry, where mass elements at the poles emit 25\% more neutrino energy than those at the equator, the ISW-like effect would result in an $\alpha \sim -0.02$ neutrino energy flux asymmetry, where the SMS emits roughly 2\% less neutrino flux in the polar direction than the equatorial direction. This assumed number for rotation-induced temperature anisotropy is chosen for illustrative purposes only, with only the proviso that the rotation speeds be so modest as to not alter the collapse significantly.

Note that $\alpha$ and $\eta$ have opposite signs. This is because the more emissive polar regions ($\eta > 0$) will create more neutrino energy flux in the equatorial directions than polar directions. This is because the inward directed flux will be suppressed by the ISW-like effect. In addition, $\alpha$ is smaller in magnitude than $\eta$ because of an averaging effect over outward directed neutrino trajectories that reduces the size of the asymmetry.

%%%%%%%%%%%%%%%%%%%%
%%---------------- Figure 1 ------------------   
%%---------------- Figure 1 ------------------   
%%%%%%%%%%%%%%%%%%%%

Several factors stemming from the strong gravitational field and relativistic environment might also alter the neutrino energy flux asymmetry. Direction beaming effects will be most important when the matter moves at relativistic velocity. This will happen only near the BH formation point. In spite of that, the redshift will dominate over the beaming effects whenever the in-falling fluid elements are moving close to the speed of light~\cite{Cardall:1997bi}. As a result, the beaming contribution to the neutrino luminosity should not be dominant. Another factor to consider is the deflection of neutrino trajectory in the strong gravitational field regime. Again, reference~\cite{Cardall:1997bi}, studying the neutron star regime, shows that where this effect is significant, redshift is dominant. Just how significant the null trajectory-bending effect could be in altering the neutrino emission asymmetry requires a fully general relativistic simulation, which is beyond the scope of this paper.

%------------------------------------------------------------------------------------------------------------------------
%  Section 3, Gravitational waves from anisotropic neutrino emission
%------------------------------------------------------------------------------------------------------------------------
\section{Gravitational waves from anisotropic neutrino emission}\label{quadrupole_approximation}
Anisotropic neutrino energy transport and emission can radiate gravitational waves so long as there is a time-changing quadrupole moment in the neutrino flux. This type of gravitational wave signal was first analyzed by Epstein~\cite{Epstein:1978dv}. Since then the formalism has been applied to core-collapse supernovae in several studies~\cite{Burrows:1995bb, Mueller:1997, Fryer:2004wi, Ott:2006qp, Dessart:2006gd, Kotake:2006aq}.

In this paper, we use the same formalism but deal with a completely different object and environment. SMSs have mass density several orders of magnitude lower than the density of core-collapse supernovae. Even at the onset of black hole formation, the density in the center of the SMSs we consider is no more than $10^9\,{\rm g}\,{\rm cm}^{-3}$, while the density of core-collapse supernovae reaches nuclear matter density, $\sim 10^{14}\,{\rm g}\,{\rm cm}^{-3}$, or higher. In core-collapse supernovae anisotropy in the neutrino emission and outgoing neutrino flux stems from inhomogeneity on the surface of the neutrino-sphere, roughly the proto-neutron star surface. By contrast, as discussed in the last section, in the SMS case anisotropy in the neutrino emission and outgoing flux is produced by temperature anisotropy in the homologous core.

%%--------------------------------
%%     Section 3A
%%--------------------------------
\subsection{Collapsing SMSs and gravitational radiation}\label{strain_amplitude}

Consider the collapse of a SMS that anisotropically emits a burst of neutrinos with total energy $E_{\nu,~\rm loss}$ over a burst time scale $\Delta t$. The gravitational wave strain measured distance $d$ away from this prodigious neutrino burst can be estimated with the quadrupole moment approximation, $h \approx 2 \ddot{I} / d$. If $\alpha$ represents the polar-equatorial neutrino emission asymmetry, the neutrino mass-energy density will have an asymmetric component $\alpha E_{\nu,~\rm loss} / (4 \pi R^2 \Delta t)$, where $R$ is the radius of the homologous core. This corresponds to a quadrupole moment $I = \alpha E_{\nu,~\rm loss} R^3 / (15 \Delta t)$.

The characteristic neutrino burst time is the dynamical timescale of the collapsing homologous core, which is approximately the light crossing time across the homologous core near BH formation, $\Delta t \approx 2M^{\rm HC}$. This is also roughly the free-fall time near BH formation. Because of the steep temperature dependence of $e^\pm$-pair annihilation neutrino energy emissivity, most neutrinos are radiated close to BH formation. Consequently, we take $R$ as the Schwarzschild radius of the homologous core, $2 M^{\rm HC}$. Assuming the total energy release in neutrinos is a fraction $\beta$ of the homologous core rest mass, the gravitational wave strain can be estimated as
\begin{equation}
	h \approx 6.5\times 10^{-20} {\alpha\beta} \left(\frac{M^\text{HC}}{10^5\,M_\odot}\right) \left(\frac{10~{\rm Gpc}}{d}\right).
	\label{eq:strain_estimate}
\end{equation}
Note that cosmological redshift will increase the burst duration at the detector. The neutrino burst time $\Delta t$ in the source's rest frame (including SMS gravitational redshift effects) will be redshifted to $\Delta t_m = \Delta t \left(1+z\right)$ in the detector's rest frame. Table~\ref{table:GWcharacteristics} shows the characteristics of gravitational wave signals from collapsing SMSs at redshift $z = 7$ and with a $2\%$ neutrino emission asymmetry. This table presents these estimates for two different calculations of $E_{\nu,~\rm loss}$.

Of course, the strain derived by using Eq.~(\ref{eq:strain_estimate}) is only an order-of-magnitude estimate. One flaw in this estimate is the approximation of the time-derivative as the inverse of the characteristic neutrino burst time, which would imply a single-frequency wave. However, because the neutrinos emitted during SMS collapse are gravitationally unbound, the gravitational wave generated by the neutrino burst is a GWM with broad-band characteristics. To get the correct power spectrum of the gravitational radiation, one should include Fourier components at all frequencies. Nevertheless, Eq.~(\ref{eq:strain_estimate}) serves to capture the GWM strain amplitude to be expected from the time-changing energy flux and quadrupole moment of the neutrino field associated with SMS collapse.

%%%%%%%%%%%%%%%%%%%%%%%%%%%%%%%%%%%%%%%%%%%%%%%%
\begin{table}[t]
	\caption{Gravitational waves from collapse of SMS at redshift 7 with a $2\%$ neutrino emission asymmetry.} 
	\centering 
	\begin{ruledtabular}
	\begin{tabular}{c  c c c c} % centered columns (4 columns)
		 $M^{\rm HC}=1\times10^5\,M_\odot$ & Shi \& Fuller & ~Linke et al.  \\ [0.5ex] 
			\hline 
			\\
			$E_{\nu,~\rm loss}$ & $3.6\times 10^{57}~{\rm erg}$ & $1.1\times10^{56}~{\rm erg}$ \\
			Fraction of rest mass $\beta$ & $2\times 10^{-2}$ & $5\times 10^{-4}$\\
			GW strain $h$ & $3.0\times10^{-23}$ & $8.3\times10^{-25}$ \\[1ex] 
	\end{tabular}
	\end{ruledtabular}
	\label{table:GWcharacteristics} % is used to refer this table in the text
\end{table}
%%%%%%%%%%%%%%%%%%%%%%%%%%%%%%%%%%%%%%%%%%%%%%%%

%%--------------------------------
%%     Section 3B
%%--------------------------------
\subsection{Neutrino burst-generated gravitational waves with memory}\label{waveform}

We follow the formalism in Ref.~\cite{Epstein:1978dv}  to calculate the key features of the form of the gravitational wave with memory (GWM) generated by neutrino emission in SMS collapse. These results also can be derived via a time-changing quadruple moment approach, as detailed in Appendix~\ref{appendix a}. The gravitational wave measured at time $t$ by an observer at distance $d$ from the SMS source was generated by that source at retarded time $t' = t-d$. The corresponding dimensionless gravitational wave strain is~\cite{Epstein:1978dv, Mueller:1997}
	\begin{equation}
	\begin{aligned}
		\Delta h^{\rm TT}_{+} + \Delta h^{\rm TT}_{\cross}  = \frac{2}{d} &\int_{-\infty}^{t-d} L_\nu\left(t'\right) dt' \\ 
			&\int {F\left(t', \Omega'\right)}\left( 1+ \cos{\theta}  \right)e^{i2\phi}  \: d\Omega'\: 
		\label{GWstrain_waveform}
	\end{aligned}
	\end{equation}
where $L_\nu\left(t'\right)$ is the neutrino energy luminosity at the retarded time, $F\left(t', \Omega' \right)$ is the emission angular distribution function and $d\Omega' = \sin\theta^\prime\,d\theta^\prime\, d\varphi^\prime$ is the solid angle enclosing the source. The superscript TT denotes the transverse traceless gauge and \lq\lq strain\rq\rq\ is the metric deviation, which is identical to the trace reverse in this gauge. Here we introduce the detector's (observer's) frame ${\it xyz}$ and the source frame ${\it x'y'z'}$, as shown in Fig.~\ref{fig:coordinate} in Appendix~\ref{appendix a} \---  the detector is at a distant location $d$ along the observer's $z$-axis in this figure. With the orientation of axes in this figure, the two gravitational wave polarizations at the detector are $h_+^{\rm TT} \equiv h_{xx}^{\rm TT}=-h_{yy}^{\rm TT}$ and $h_{\cross}^{\rm TT} \equiv h_{xy}^{\rm TT} =h_{yx}^{\rm TT}$.

To simplify the calculation, we take the emission angular distribution to be time-independent and axisymmetric about the $z'$ axis:
	\begin{equation}
		F\left(\Omega'\right) = \frac{1+\alpha \cos^2{\theta'}}{4\pi\left( 1+\alpha/3 \right)}.
		\label{eq:anuglar_distribution}
	\end{equation}
The angular distribution of neutrino emission is enhanced at the two poles relative to the equator when $\alpha>0$, and in the equatorial plane relative to the poles when $\alpha<0$. The scenario that we describe in Sec.~\ref{Anisotropic nu emission} has $\alpha<0$. Because of the $\phi'$-independence of the emission distribution in Eq.~(\ref{eq:anuglar_distribution}), it can be shown that the only relevant polarization in Eq.~(\ref{GWstrain_waveform}) is \lq\lq plus\rq\rq\ polarization, $h_+^{\rm TT} = h_{xx}^{\rm TT} = - h_{yy}^{\rm TT}$.

After integration over all solid angles in Eq.~(\ref{GWstrain_waveform}), the gravitational wave strain is:
	\begin{equation}
		\Delta h^{\rm TT}_{+} = \Delta h_{xx}^{\rm TT} = - \Delta h_{yy}^{\rm TT}  = \frac{E_{\nu~{\rm loss}}}{d} \times \frac{\alpha \sin^2{\xi}}{3+\alpha}.
		\label{eq:after_integration}
	\end{equation}
As expected, the gravitational wave strain is zero when the detector is located along the polar axis of the source ($\xi=0$ or $\pi$) and maximal in magnitude when the detector is located in the source's equatorial plane ($\xi = \pi/2$).

As noted, the gravitational wave signal generated by anisotropic neutrino emission is a GWM. The \lq\lq memory\rq\rq\ effect is so named because this gravitational wave type results in a nonzero net strain after the signal has passed the detector. In other words, its passage imprints a permanent proper displacement between two freely falling masses. The GWM waveform in the time-domain would look like a DC offset signal, but with a rise time $\Delta t_m$:
	\begin{equation}
		h\left( t \right) = 
			\begin{cases} 
      				0 & t<-\Delta t_{m}, \\
      				\Delta h^{\rm TT}_{xx} \left( 1+t/\Delta t_{m} \right) & -\Delta t_{m} < t <0, \\
      				\Delta h^{\rm TT}_{xx} & t > 0,
   			\end{cases}
   		\label{eq:waveform}
	\end{equation}
where $\Delta h^{\rm TT}_{xx}$ is calculated in Eq.~(\ref{eq:after_integration}).

%-------------------------------------------------------------------------------------------------------------------------
% Section 4, Signal to noise ratio
%-------------------------------------------------------------------------------------------------------------------------
\section{The Signal To Noise Ratio}\label{signal to noise ratio}

In this section we compute the signal-to-noise ratio of the neutrino burst-generated gravitational wave signals and we consider the prospects for detecting these signals with space-based laser interferometry. The sky-averaged squared signal-to-noise ratio is $\langle\text{SNR}^2\rangle = \int_0^\infty \left[{h_c(f)}/{h_n(f)}\right]^2 {df}/{f}$. Here $h_c\left(f\right)$ is the GWM's characteristic strain at frequency $f$ and is defined as 
	\begin{equation}
		h_c\left(f\right)_{\rm mem} = 2f \langle\lvert\tilde{h}_+(f)\rvert^2\rangle^{1/2},
		\label{eq:hc}
	\end{equation}
where $\tilde{h}_+(f)$ is the Fourier transform of the GWM plus-polarization strain (metric deviation) in Eq.~(\ref{eq:waveform}):
	\begin{equation}
		\tilde{h}_+\left(f\right)  = \Delta h^{\rm TT}_{xx} \: \frac{-i e^{-\pi i f \Delta t_{m}}}{2\pi^2 f^2 \Delta t_{m}}\: \sin{\left(\pi f \Delta t_{m}\right)}.
	\label{eq:fourier_transform}
	\end{equation}
The $\langle...\rangle$ in Eq.~(\ref{eq:hc}) denotes the average over the sky position and polarization of the source, {\it i.e.}, the average over $\xi$. $h_n\left(f\right)$ is the characteristic detector noise amplitude obtained after taking the average of sky-location and polarization angle, {\it i.e.}, $h_n\left(f\right) = {\sqrt{fS_n\left(f\right)}}/ { \langle F_{+}^{2}(\theta,\phi,\psi)\rangle^{1/2} }$, where $S_n\left(f\right)$ is the detector's one-sided noise spectral density and $F_{+}(\theta,\phi,\psi)$ is the detector's beam pattern function. The value of $\langle F_{+}^{2} \rangle$ for detectors like DECIGO and BBO is $1/5$, and 3/20 for detectors like LISA~\cite{300yearsofgravitation, Apostolatos:1994mx, Barack:2003fp, Yagi:2011wg}.

%%%%%%%%%%%%%%%%%%%%
%%---------------- Figure 2 ------------------   
\begin{figure}[t]
\includegraphics[width=0.45\textwidth]{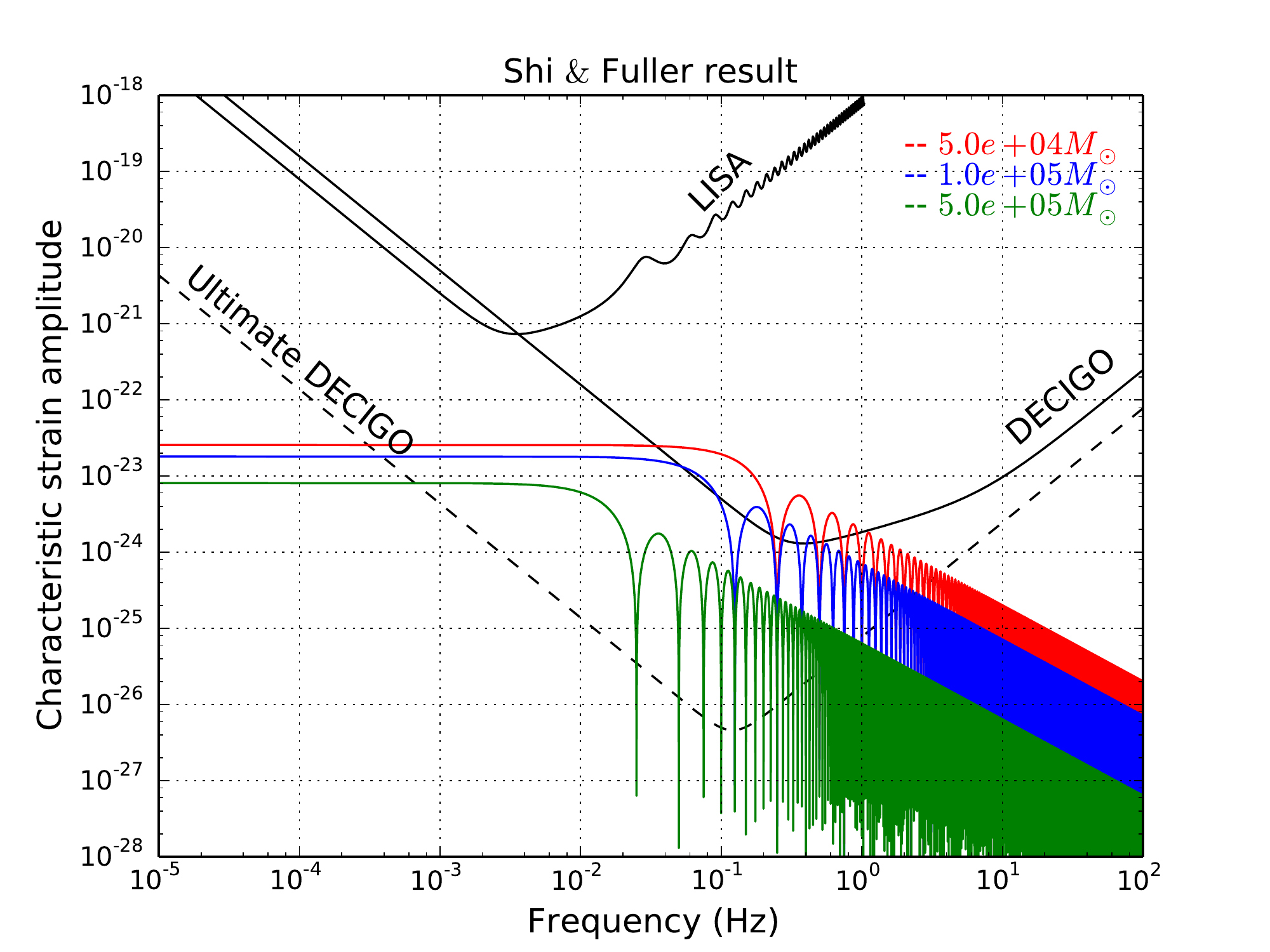}
\includegraphics[width=0.45\textwidth]{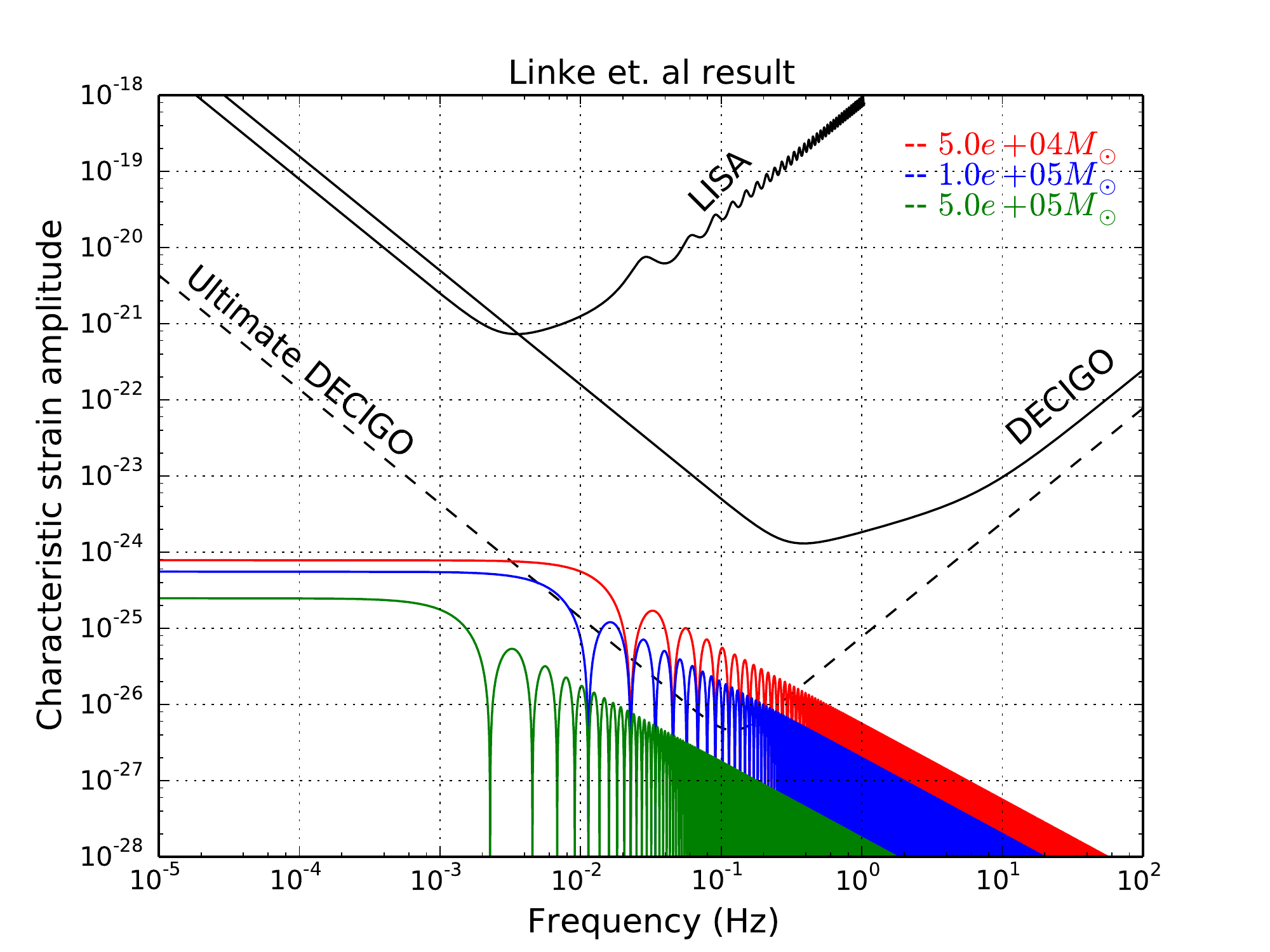}
\caption{Sky-averaged characteristic strain $h_c$ as a function of frequency from the neutrino burst-generated GWM signal accompanying the collapse of SMSs at $z=7$, with $\alpha = -0.02$ and with homologous core masses $5\times 10^4\,M_\odot$ (red curve), $10^5\,M_\odot$ (blue curve) and $5\times 10^5\,M_\odot$ (green curve), as labeled. The panels on the top and the bottom are based on the results for integrated neutrino luminosity from Shi $\&$ Fuller and Linke et al., respectively. The two black solid lines and the black dash line denote the sky-averaged noise curves for LISA, DECIGO and Ultimate DECIGO, as labeled.}
\label{fig:strain}
\end{figure}
%%---------------- Figure 2 ------------------   
%%%%%%%%%%%%%%%%%%%%

In the low frequency limit where $f \ll 1/\Delta t_m$, Eq.~(\ref{eq:fourier_transform}) becomes $\tilde{h}\left(f\right) = \Delta h/ 2\pi i f$ and the dimensionless characteristic strain $h_c$ approaches a frequency-independent value $\Delta h/\pi$ (here we use $\Delta h$ to denote generically the metric deviation signals referred to above, e.g., $\Delta h_{x x}^{\rm TT}$, etc.). This is one of the interesting properties of the gravitational wave memory effect. These low frequency characteristics of GWM are sometimes referred to as the ``zero frequency limit''~\cite{Smarr:1977fy, Turner:1978jj, Payne:1984ec}.

In general, detectors with high sensitivity at low frequency are ideal for memory-type gravitational waves detection. 
Consider, for example, the pulsar timing array (PTA)~\cite{foster1990constructing, lorimer2008binary}, which is most sensitive in the nano-Hertz frequency band. A gravitational wave memory signal in this band that can be treated as an extreme low frequency wave is potentially ``audible'' to PTA. But one important factor that limits the sensitivity of the PTA in detecting GWM signals is the resolution of the best clock in the world. The pulse arrival time shifted by gravitational waves is $\Delta t/t \sim \Delta h$; on the other hand, the stability of the best clock, which has a strain sensitivity at a level $\sim 10^{-15}$ after integrating the data for 10 years~\cite{Jenet:2011me}, is still far short of what is required to detect gravitational waves from collapsing SMSs, where we might expect $\Delta h \sim10^{-23}$.

Fortunately, space-based gravitational wave detectors, for example DECIGO and BBO~\cite{Harry:2006fi, Crowder:2005nr}, with optimal sensitivity to frequencies in the deci-Hertz band, and high peak sensitivity ($h_c\sim 10^{-24}$), could be ideal for detecting GWM from neutrino bursts from SMS collapse. Serendipitously, the SMS homologous core mass range giving the largest fraction of rest mass radiated as neutrinos also produces GWM with frequencies more or less coincident with the optimal sensitivity frequency range for DECIGO and BBO.

%%%%%%%%%%%%%%%%%%%%
%%---------------- Figure 3 ------------------   
\begin{figure}[t]
\includegraphics[width=0.45\textwidth]{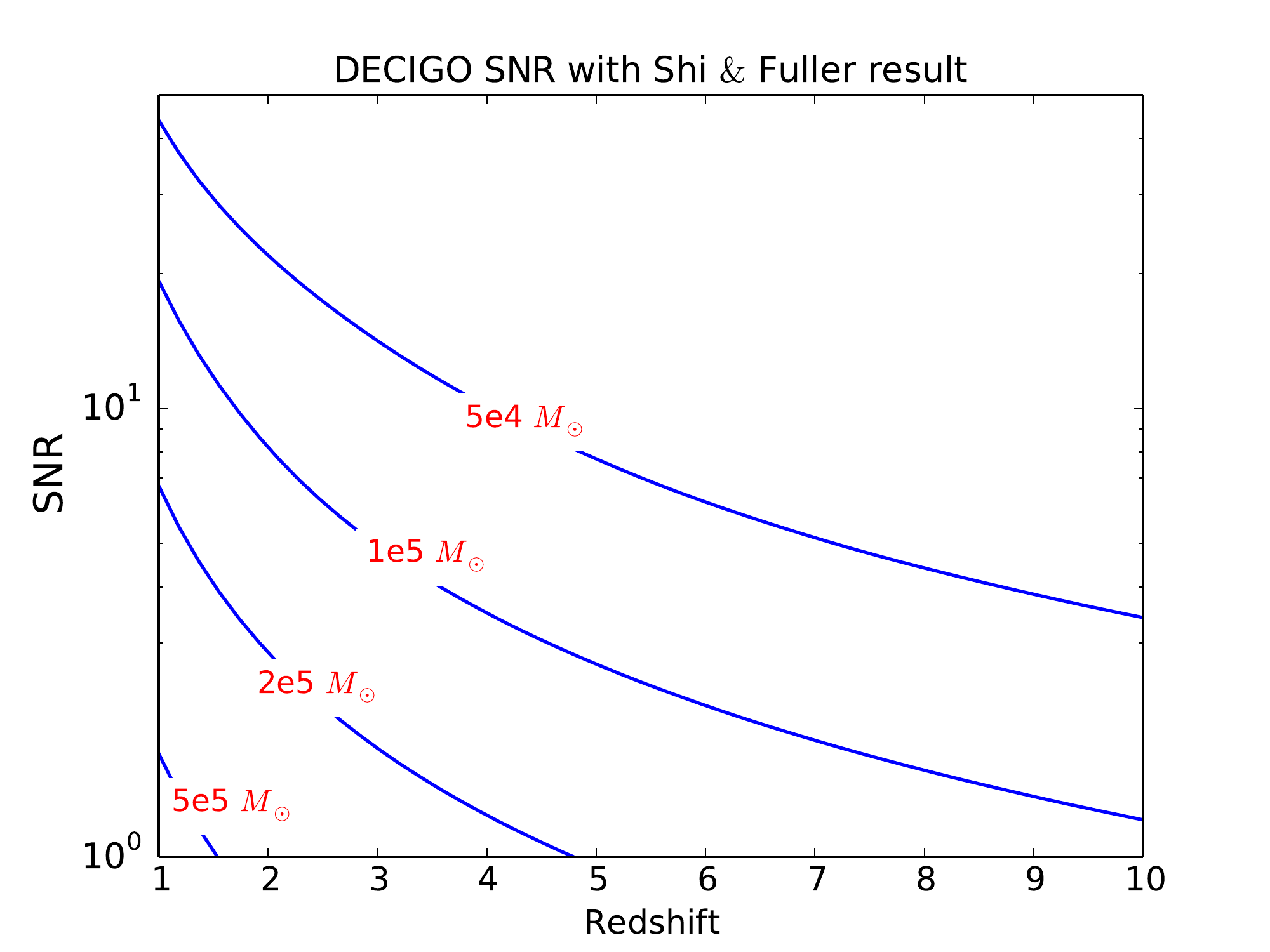}
\includegraphics[width=0.45\textwidth]{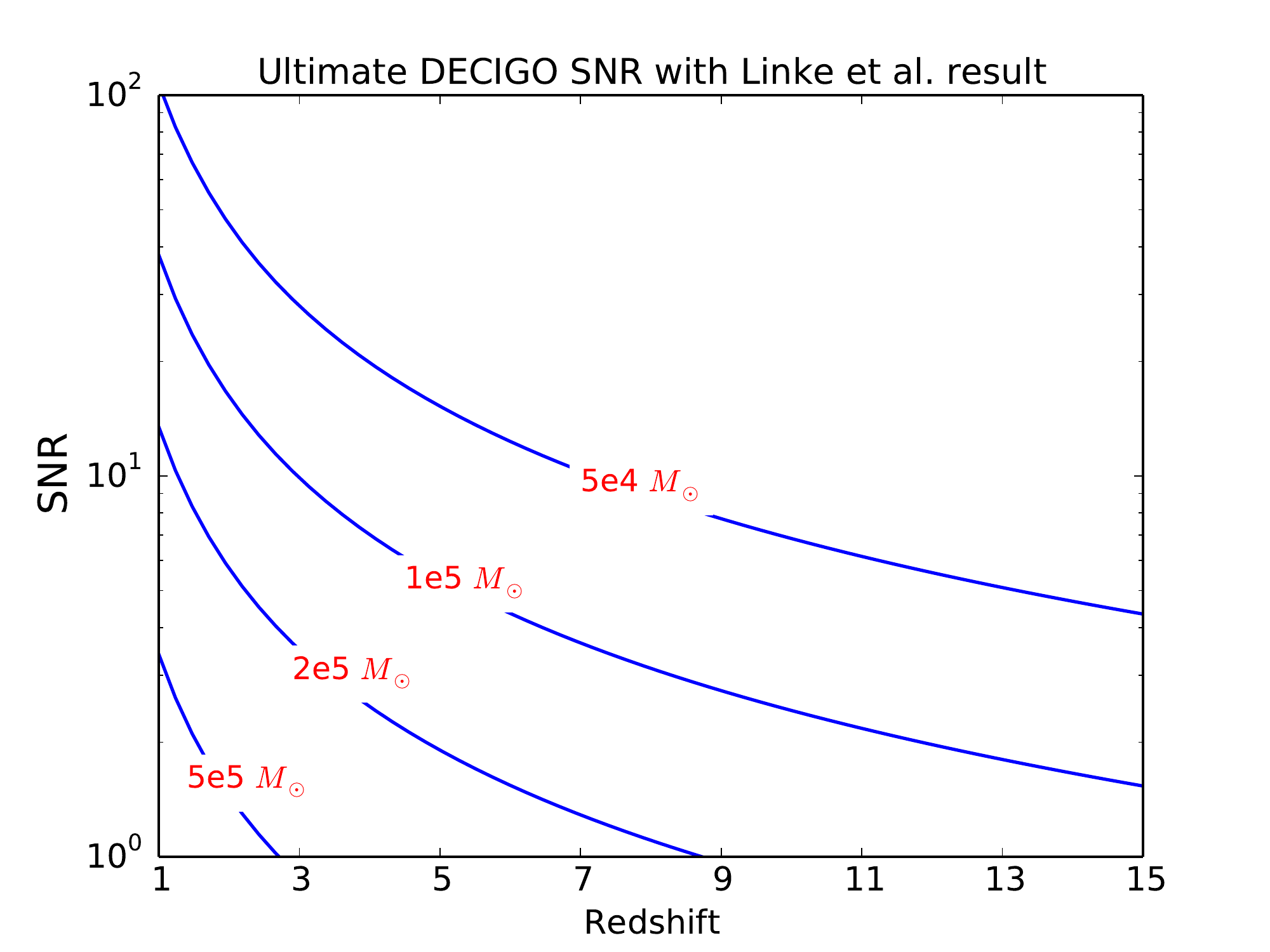}
\caption{The vertical axes in the top and bottom panels show sky-averaged signal-to-noise ratios SNR for DECIGO with the Shi $\&$ Fuller result and Ultimate DECIGO with Linke et al. result for overall neutrino burst characteristics, respectively. Here we take asymmetry parameter $\alpha = -0.02$. Each contour line denotes the final homologous core mass of a collapsing SMS, as labeled.}
\label{fig:SNR}
\end{figure}
%%---------------- Figure 3 ------------------ 
%%%%%%%%%%%%%%%%%%%%  
%%
%%%%%%%%%%%%%%%%%%%%
%%---------------- Figure 4 ------------------   
\begin{figure}[t]
\includegraphics[width=0.5\textwidth]{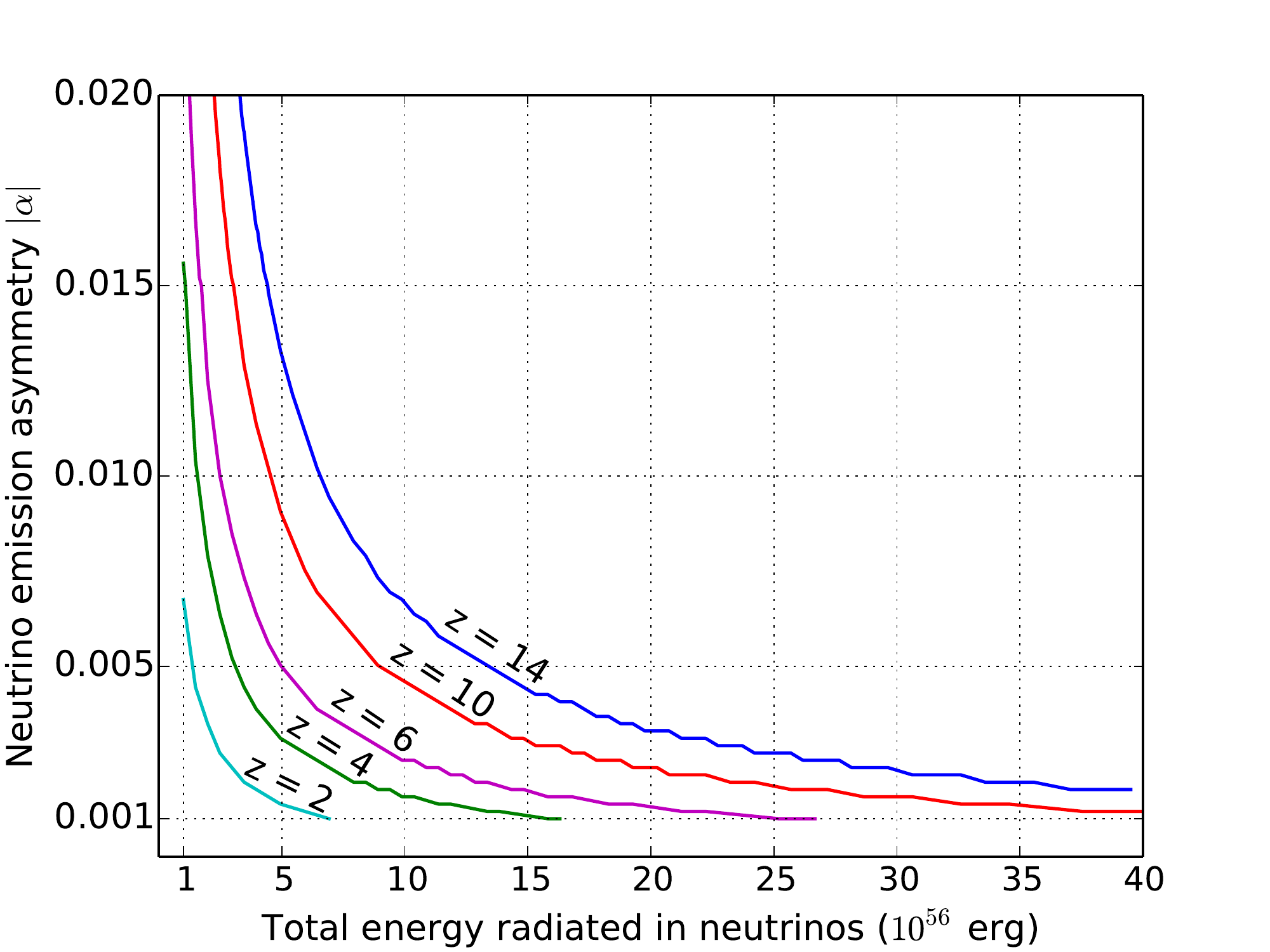}
\caption{Contours of detectability redshift $z$ (as labeled) as a function of the neutrino emission asymmetry and the total energy radiated in neutrinos. These results were calculated assuming an SMS homologous core mass $M^{\rm HC} = 10^5\,M_\odot$ with neutrino burst time calculated with the Linke et al. result.
Each contour curve shows the redshift for which SNR = 5 for Ultimate DECIGO.} 
\label{fig:Asymmetry_vs_Energy}
\end{figure}
%%---------------- Figure 4 ------------------   
%%%%%%%%%%%%%%%%%%%%

Fig.~\ref{fig:strain} shows a comparison between the Shi \& Fuller and Linke et al. integrated neutrino luminosity results for the sky-averaged characteristic strains of the SMS neutrino burst-generated GWM signals estimated here. LISA is a gravitational wave interferometry antenna in Earth-like solar orbit with arm length 2.5 Gm~\cite{Audley:2017drz}. With the currently envisioned LISA design sensitivity, the gravitational wave signal from the collapse of SMSs at $z\gtrsim0.1$ is too weak to be detected, for both results for overall neutrino emission.

DECIGO is also a gravitational wave interferometry antenna in Earth-like solar orbit, but with a 1000 km arm length and covering the mid-frequency ($\sim$ 0.1 Hz) gravitational wave band with $h_{\rm rms}\sim2\times10^{-24}$~\cite{Seto:2001qf}. Its high sensitivity at $f\sim0.1~{\rm Hz}$ is ideal for the detection of SMS neutrino burst-generated gravitational wave signals which have characteristic timescale $\sim$ 1~s to 10~s. With the Shi \& Fuller result, a GWM signal for an SMS with $M^{\rm HC} \approx 5\times10^4\,M_\odot$ will be ``audible'' with DECIGO (here assuming a basic ``set''~\cite{Yagi:2011wg} of detectors) at $z=7$, and much higher redshifts for envisioned ultimate DECIGO design parameters~\cite{Seto:2001qf}. The Linke et al. result has about a factor of 30 lower total neutrino energy release and a factor of 11 longer neutrino emission time than the Shi \& Fuller result. These differences imply a reduction in strain amplitude and lower signal frequencies relative to results of calculations carried out with the Shi \& Fuller estimates. With the Linke et al result, a GWM signal for an SMS with $M^{\rm HC} \approx 5\times10^4\,M_\odot$ is too faint to be seen at $z\gtrsim0.1$ with the basic DECIGO configuration, but will be detectable with Ultimate DECIGO at $z =7$.

Fig.~\ref{fig:SNR} gives examples of the expected SNR for our estimated SMS neutrino burst-generated GWM as a function of redshift for a range (contours) of homologous core masses. The results shown in this figure use both the Shi \& Fuller (top panel) and Linke et al. (lower panel) estimates for overall neutrino burst luminosity. In each of these example calculations we take the neutrino energy flux asymmetry parameter to be $\alpha = -0.02$. This figure provides insight into the prospects for detection of these GWM signals. We show (upper panel) the most optimistic estimate of neutrino burst luminosity and most favorable (highest) rest frame frequency paired with the least sensitive version of DECIGO, and the least favorable estimate of neutrino emission and rest frame frequency range paired with the most sensitive and capable version of DECIGO planned, i.e., \lq\lq Ultimate\rq\rq\ DECIGO. Based on Shi \& Fuller result, the GWM signal for an SMS with $M^{\rm HC} =5\times10^4\,M_\odot$ and $\alpha = -0.02$ could be detected by DECIGO with SNR $> 5$ out to redshift 7.
With the same mass and asymmetry, the GWM signal with Linke et al. result is not detectable with basic DECIGO, but is detectable by Ultimate DECIGO with SNR $> 5$ out to redshift 13.

Fig.~\ref{fig:Asymmetry_vs_Energy} provides insight into detectability of neutrino-burst-generated GWM. In this figure we show contours of detectability redshift as a function of neutrino emission asymmetry and the total energy radiated in neutrinos for a SMS with homologous core mass $M^{\rm HC} = 10^5\,M_\odot$. Here the contours of redshift  \lq\lq detectability\rq\rq\ indicate a SNR$\ge 5$ in Ultimate DECIGO.
The total neutrino emission for this particular example is $3.6 \times10^{57}~{\rm ergs}$, as calculated with the Shi \& Fuller neutrino emission result and approximately $10^{56}~{\rm ergs}$ with Linke et al. result.
All of these estimates are intriguing, suggesting that deci-Hertz gravitational wave detectors may be able to probe massive black hole production and associated physics at redshifts at, and even well beyond, those of the epoch of re-ionization.

SMS collapse events with a given mass and given neutrino emission asymmetry could be detected to even larger distances if nuclear burning  prior to, or during, collapse causes the entropy to increase, in turn causing a larger fraction of SMS rest mass to be radiated as neutrinos. But nuclear burning and rotation also can decrease the chances for detection of SMS neutrino bursts.

In this vein, we should emphasize that all of our estimates are rough, and many issues in SMS physics remain open, as discussed above in Sec.~\ref{Characteristics_SMS}. For example, the neutrino emission calculations in both Shi \& Fuller and Linke et al. results do not include the possible effects of nuclear burning on the SMS's collapse adiabat, nor do they incorporate the phasing of this nuclear energy input with the post-collapse build-up of infall kinetic energy in the homologous core. Moreover, if significant pressure or centrifugal support resists the free fall of the homologous core, more neutrinos can be emitted, as there is more time for emission before the formation of a trapped surface. As a consequence of this effect, however, peak neutrino emission will be shifted to a lower frequency because the collapse time will be longer than the free-fall time near the BH formation point. As illustrated by the examples in Fig.~\ref{fig:strain}, shifting the frequency of the neutrino burst-generated GWM to the low side of the DECIGO peak sensitivity frequency range impairs that detector's ability to \lq\lq see\rq\rq\ these signals at the higher redshifts.

On the other hand, the rapid release of nuclear binding energy may destroy the star in an explosion instead of forming a large remnant BH. Of course, this results in much less total neutrino emission. The calculation reported in Ref.~\cite{Chen:2014yea} suggests a possible narrow SMS mass window, centered around $M_{\rm SMS}\approx 5.5\times 10^4\,M_\odot$, where a non-rotating, primordial (zero) metallicity SMS could experience rapid, \lq\lq explosive\rq\rq\ helium burning beginning just after the conclusion of hydrogen burning and in close coincidence with the the post-Newtonian instability point. This could result in thermonuclear explosion, as not much infall kinetic energy will have been built up prior to the helium burning energy injection. Moreover, the coincidence of the triple-alpha ignition point and the onset of instability is likely what limits the SMS mass range for this behavior and targets the lower masses in the range of masses considered here -- only the lower end our mass range would have a stable main sequence.

SMS produced at later epochs, or in scenarios involving tidal disruption or stellar coalescence in a dense star cluster, may have small but non-zero initial metallicity. These could also experience thermonuclear explosion rather than collapse to a BH.  A small initial carbon, nitrogen, or oxygen (CNO) content could facilitate hydrogen hydrogen burning via the CNO cycle, and thereby allow an {\it early} break-out into the rp-process. In turn, this break-out would result in a greater rate at which nuclear energy is added as compared to that in the proton-proton hydrogen burning regime characterizing the early stages of collapse in initially zero metallicity SMSs. Nuclear energy addition {\it immediately} after the post-Newtonian instability point, before the build-up of an infall kinetic energy \lq\lq debt,\rq\rq\ enhances the chances for thermonuclear explosion.

Rotation can also enhance these chances. The study in Ref.~\cite{Montero:2011ps} shows that a rotating SMS with a mass $M_{\rm SMS}\approx 5 \times 10^5\,M_\odot$ at initial angular speed $\gtrsim 2.5 \times 10^{-5}~{\rm rad}\,{\rm s}^{-1}$ reduces the metallicity threshold for thermonuclear explosion to $Z_{\rm CNO}\approx 0.001$. Their result for a star that explodes this way shows a decrease of 10 orders of magnitude in total neutrino loss rate relative to a model that collapses to a BH. In any case, post-instability thermonuclear explosion of an SMS will decrease the total energy radiated in neutrinos and, at the same time, increase the neutrino burst timescale relative to that of a SMS that collapses to a BH. These features decrease the prospects for detecting a neutrino burst-generated GWM signal.

%------------------------------------------------------------------------------------------------------------------------
% Section 5, Other possible GWM
%------------------------------------------------------------------------------------------------------------------------
\section{Other possible GWM sources}\label{other_sources}

There are several other astrophysical sources which could produce a linear memory GWM signal with strain magnitude and overall timescale similar to those originating from a neutrino burst associated with a high-redshift SMS collapse. If a space-based laser interferometer gravitational wave observatory were to record a signal with characteristics along the lines of what we discuss above, how would we know it was actually a SMS collapse-generated GWM? Direct neutrino detection could constitute a confirmation, as the time dependence of the neutrino signal in principle could tag the event as having a SMS collapse origin \cite{Fuller:1997em, Shi:1998jx}. However, the neutrino radiation from a high-redshift SMS collapse will be difficult to detect for redshift $z \gtrsim 0.2$, though below this redshift the SMS collapse time template may allow IceCube to extract this signal \cite{Shi:1998jx}.

Detection of the gravitational-wave ringdown signal associated with the black hole produced in SMS collapse might be another way to tag the linear memory GWM signal as having a SMS collapse origin. In fact, the ringdown signal should follow neutrino burst-driven GWM by no more than one dynamical timescale, $\sim M_5^{\rm HC}$~s. However, a slowly rotating or a non-rotating SMS might not generate a gravitational-wave ringdown signal of high enough amplitude to be detected by the existing or proposed laser interferometers, especially if the SMS is at high redshift.

A non-detection of the gravitational-wave ringdown signal will force us to examine other possible sources, for example, conventional core collapse supernovae and hyperbolic binaries occurring in the local galactic group. A typical core collapse event radiates roughly $\sim 10^{53}$~erg in neutrinos in a few neutrino diffusion timescales, $\lesssim 10$~s. A supernova event occurring in the Andromeda galaxy, approximately 1~Mpc away from earth, could produce a neutrino burst-generated GWM signal with strain $\sim 10^{-23}$ on a timescale $\sim$ 1~s to 10~s, similar to the characteristics of neutrino-burst generated GWM from high-redshift SMS collapse.

Fortunately, several other counterpart signals would be expected to accompany the supernovae GWM signal, for example, the strong gravitational-wave burst without memory from the motion of the baryonic component in the source, the electromagnetic (EM) radiation, and the burst of neutrinos. The latter may be problematic to detect if the source is at an appreciable distance outside the Galaxy. The detection of any of these  counterparts could help to distinguish a GWM signal from local group core collapse supernova events and the GWM signal from a high-redshift SMS collapse. EM transients associated with conventional compact object sources should be detectable in most circumstances where their linear memory signals might be confused with those discussed here. Indeed, it is interesting to speculate on whether the EM signal from SMS collapse or explosion at high redshift might be detectable \---  the future prospects for such a detection are encouraging given the revolution occurring in time domain/transient astronomy across the EM spectrum.

There can be other compact object sources of linear memory GWM signals. Among these are hyperbolic binaries, i.e., two stars in an unbound orbit, in essence \lq\lq bremss-ing\rq\rq\ off gravitational radiation. Two stars undergoing a close, but unbound encounter, can radiate GWM signals with strain magnitude $4\, m_{\rm A}\, m_{\rm B} /(b\, r)$ on the characteristic timescale $b/v$~\cite{Kovacs:1978eu}. Here $m_{\rm A}$ and $m_{\rm B}$ are the masses of the two objects, $b$ is the impact parameter, $r$ is the distance from the observer to the hyperbolic binary source, and $v$ is the relative velocity between the two objects at closest approach. Consider two neutron stars in the Andromeda galaxy (assumed 1~Mpc distant from earth for this example) flying past to each other with a relative speed $\sim 1000~{\rm km}\,{\rm s}^{-1}$ and an impact parameter $\sim 10^4\,{\rm km}$. The gravitational wave strain from this event would be $\sim 6 \times 10^{-23}$, and the timescale over which the amplitude of this gravitational radiation is appreciable is $\sim 10$~s. We would expect no significant EM or neutrino signatures from such an event.

However, the polarization of the expected gravitational radiation from a hyperbolic encounter will be different from the polarization in SMS neutrino burst-generated GWM. As shown in Ref.~\cite{Kovacs:1978eu}, the gravitational wave generated in the hyperbolic binary encounter have both linear and circular polarization, whereas the neutrino burst-generated GWM discussed in this work would have only linear polarization. A detection of a circularly polarized component of the GWM generally would indicate that the signal was not produced in the SMS scenario discussed here. But there is a loophole. Note that at some detector inclinations relative to the orbital plane of the hyperbolic binary, the observer would receive only the linearly polarized GWM component and not the circularly polarized one (see Fig.~2 in Ref.~\cite{Kovacs:1978eu}). In that case, we would not be able to distinguish between the GWM signal coming from a hyperbolic binary and the GWM signal coming from high-redshift SMS collapse.

%------------------------------------------------------------------------------------------------------------------------
% Section 6, Conclusion
%------------------------------------------------------------------------------------------------------------------------
\section{Conclusion}

In this paper we point out an intriguing connection between neutrino burst-generated gravitational waves from the collapse of high entropy, fully convective SMSs at high redshift and the capabilities of proposed space-based gravitational wave observatories like BBO and DECIGO to detect linear memory gravitational wave signals with high sensitivity. We have made simple estimates of the expected linear memory gravitational waves (GWM) likely to be produced by SMS collapse-generated neutrino bursts and the response of these detectors to these signals. We conclude that detection of these GWM is possible in some cases and for some DECIGO detector configurations, even from SMS collapse at high redshift. Detections along these lines would open a new window on an old, but otherwise mysterious issue in relativistic astrophysics: the origin of supermassive black holes.

In the scenarios we examined, gravitational collapse of high entropy SMSs engineers prodigious neutrino production which, in turn, gives rise to a relatively unique gravitational wave signal, the GWM. The high entropy attendant to a hydrostatic SMS implies that these objects possess copious electron/positron pairs in electromagnetic equilibrium. This, coupled with the strong temperature dependence ($\propto T^9$)  of $e^\pm$-pair annihilation into escaping neutrino pairs of all flavors, guarantees that SMS collapse constitutes an prodigious engine for neutrino production.

In fact, SMSs with homologous core masses in the range $5\times{10}^4\,M_\odot$ to $5\times{10}^5\,M_\odot$ will radiate an optimal fraction of their rest mass in a burst of neutrinos, mostly produced close to the black hole formation point because of the $T^9$ neutrino emissivity dependence. Neutrinos from lower mass stars will likely suffer scattering-induced trapping, cutting down the amplitudes and decreasing the frequency of gravitational waves produced, while higher mass SMSs do not get hot enough to radiate a significant fraction of their mass in neutrinos before black hole formation. The collapse of a $10^5\,M_\odot$ SMS is likely accompanied by a few percent of its gravitational binding energy being radiated as neutrinos, on a timescale $\sim$ 1~s to 10~s. An asymmetry in the outgoing neutrino energy flux can create a characteristic GWM signal, observable in the frequency bands where DECIGO and BBO are most sensitive.

For example, a modest rotation of the SMS could result in a small temperature and neutrino emission asymmetry. In an otherwise static and neutrino-transparent SMS, this would not produce an appreciable quadrupole moment in the neutrino field. However, in a non-static, collapsing SMS, the neutrino direction symmetry is broken, and a differential blueshift-redshift effect, much like the integrated Sachs-Wolfe (ISW) effect for photons propagating through evolving density fluctuations/potential wells in the early universe, serves to imprint any temperature asymmetry or inhomogeneity on the outgoing neutrino energy flux \--- this can give a time-changing quadruple moment in the neutrino mass-energy field and, hence,  gravitational radiation. Using Shi \& Fuller's result for neutrino energy luminosity, the neutrino burst-generated GWM signal produced from the collapse of an $M^{\rm HC} = 5\times 10^4\,M_\odot$ SMS could be observed with ${\rm SNR} > 5$ for DECIGO out to redshift 7 and for Ultimate DECIGO out to redshifts of order $\sim 100$. Using Linke et al.'s result for the neutrino energy luminosity, they would be observable with out to redshift $z \sim 13$ with ${\rm SNR} > 5$ in  the Ultimate DECIGO configuration. The unique characteristics of the DECIGO detector response to a linear memory gravitational wave should allow this detector to tag this signal as a GWM.

There are many pitfalls and unresolved issues in our estimates. We have discussed several of these, including the effects of nuclear burning and the phasing of this energy input with hydrostatic SMS evolution, collapse, and neutrino emission. Near BH formation, the collapse timescale \-- over which most of the neutrinos are emitted \-- may be significantly larger than the free-fall timescale we have employed in our calculations. This could shift the timescale of the GWM longer and out of the most sensitive frequency range of the detectors like BBO/DECIGO.

Hydrodynamic evolution itself could be impacted by the competing processes of nuclear burning and neutrino energy loss. For example, radiation pressure will resist the infall of the homologous core, resulting in a collapse timescale larger than the free-fall timescale. On the other hand, a more extended collapse time will increase the integrated neutrino emission, and therefore increase the GWM strain. The coupled nuclear, weak interaction, rotation, and hydrodynamic evolution of SMS stars remains a fascinating, if complicated story. How these issues play out in detail could affect the GWM estimates we make here. Obviously, a key conclusion of our work here is that more sophisticated calculations including these and other effects are warranted.

It remains an open question whether high entropy, fully convective SMSs form at high redshifts, and if they do form, whether the BHs they produce are the seeds for the formation of high redshift SMBHs. For the purposes of this study, we are agnostic on these issues. However, the detection of GWM signals attributable to the neutrino burst from these high redshift SMSs may provide an intriguing hint toward solving, or narrowing, the problem of the formation of SMBHs in the high redshift universe. It is remarkable that the envisioned space-based gravitational wave observatories like BBO/DECIGO could be poised to probe this physics in a nearly unique way.

%------------------------------------------------------------------------------------------------------------------------
% Acknowledgments
%------------------------------------------------------------------------------------------------------------------------
\begin{acknowledgments}
We thank Chris Fryer, Lee Lindblom, Wei-Tou Ni, S. Tawa, and Massimo Tinto for valuable discussions. This work was supported in part by National Science Foundation Grants No. PHY-1307372 and No. PHY-1614864. We also would like to acknowledge a grant from University of California Office of the President.
\end{acknowledgments}

%------------------------------------------------------------------------------------------------------------------------
% Appendix  A
%------------------------------------------------------------------------------------------------------------------------
\appendix
\section{Quadrupole Moment Approximation}\label{appendix a}
Epstein~\cite{Epstein:1978dv} has given a rigorous derivation of gravitational radiation generated from a neutrino burst via direct integration of the {\it linearized} inhomogeneous Einstein field equations. In the following, we derive this result in the weak-field quadrupole moment approximation.

We can break up the neutrino burst into N components and label them by index $\alpha$ = 1, 2, 3, ..., N. The mass-energy density distribution can be written in the point-mass description:
	\begin{equation}
		\rho_{\nu}\left(t_r, {\bf x'}\right) = \sum_{\alpha} \:\frac{M_\alpha}{\sqrt{1-v_\alpha^2}}\:\delta \left( {\bf x'} - {\bf r}_\alpha \right),
	\end{equation}
where $M_\alpha$, $v_\alpha$ and ${\bf r}_\alpha$ are the rest mass, velocity and position of the ${\alpha}$th neutrino.

Assuming constant neutrino velocity, the second time-derivative of the mass quadrupole moment tensor becomes
		\begin{equation}
			\begin{split}
				 \ddot{I}^{jk}\left(t_r\right) &= 2 \;\sum_\alpha \:\frac{M_\alpha}{\sqrt{1-v_\alpha^2}}\: v_\alpha{}^{j} \: v_\alpha{}^k\\
				&= 2\int \rho_\nu\left(t_r,x'\right)  \: \frac{n^j n^k}{1-{\vec{\bf N}}\cdot \hat{\bf n} }\: d^3x'
			\label{eq:second_time_derivative}
			\end{split}
		\end{equation}
where $\vec{\bf N}$ is the unit vector between the source and the detector and $\hat{\bf n}$ is the unit vector of the neutrino flux directed into the solid angle $d\Omega'$. 
The second step of Eq.~(\ref{eq:second_time_derivative}) assumes neutrinos travel at the speed of light and we interpret the ${M_\alpha}/{\sqrt{1-v_\alpha^2}}$ to be the {$\alpha$}th neutrino's energy measured in the detector's rest frame. 
The $(1-{\vec{\bf N}}\cdot \hat{\bf n})^{-1}$ term comes from the Lienard--Wiechert solution. Now apply the gravitational wave quadrupole formula and obtain
		\begin{equation}
			\begin{split}
				\Delta h^{jk}_\text{TT}\left(t, {\bf x}\right) = \frac{4}{d} \: \int_{-\infty}^{t-d} \int \frac{d^2E_\nu}{dt'd\Omega'}  \: \Bigg[ \frac{n^j n^k}{1- {\vec{\bf N}} \cdot \hat{ \bf n}} \Bigg]^\text{TT} \: d\Omega' \:dt'.
			\end{split}
			\label{eq:strain_of_memory}
		\end{equation}
Here $d$ is the distance from the source to detector; $d\Omega'$ is the solid angle enclosing the source; $E_\nu$ is the total energy emitted as neutrinos. Evaluating Eq.~(\ref{eq:strain_of_memory}) yields gravitational waves with linear memory (GWM) from a burst of neutrinos: gravitational wave strain $h^{jk}_\text{TT} =0$ before the GWM arrives and accumulates to a nonzero value $\Delta h^{jk}_\text{TT}$ after the gravitational wave passes the detectors.

If the emission has spherical symmetry, then there is no gravitational signature -- this is Birkhoff's theorem; but if there is a small anisotropy in the neutrino emission $dE_\nu/d\Omega'$, then the integral in Eq.~(\ref{eq:strain_of_memory}) is nonzero and therefore the memory strain accumulates to a nonzero value. The function $d^2E_\nu/d\Omega'dt'$ in Eq.~(\ref{eq:strain_of_memory}) can be written as $L_\nu\left(t'\right) F\left(t', \Omega'\right)$, where $L_\nu\left(t'\right)$ is the neutrino luminosity and $F\left(t', \Omega'\right)$ is the emission angular distribution with $\int F\left(t', \Omega'\right) \: d\Omega' = 1$.

%%%%%%%%%%%%%%%%%%%%
%%---------------- Figure 5 ------------------   
\begin{figure}[tp!]
\includegraphics[width=0.8\linewidth]{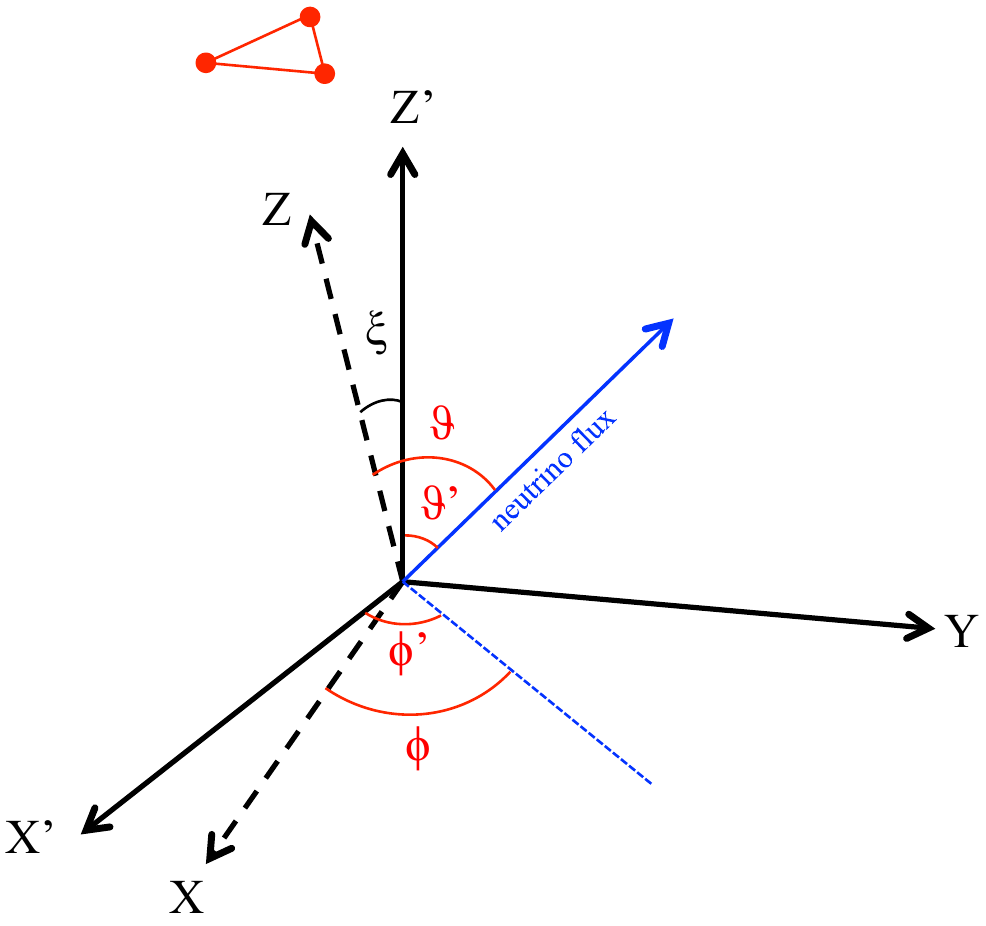}
\caption{${\it xyz}$ and ${\it x'y'z'}$ are coordinate systems of the detector and the source, respectively. 
The source is located at the origin and the detector is far out in the $\hat{z}$ direction. 
We choose the neutrino emission distribution to be axisymmetric about the $z'$ axis. 
The $x-z$ and $x'-z'$ planes are coplanar and differ by a rotation by angle $\xi$ about the y axis.}
\label{fig:coordinate}
\end{figure}
%%---------------- Figure 5 ------------------   
%%%%%%%%%%%%%%%%%%%%

Placing the detector at the transverse direction of the gravitational wave, say along the $z$ axis in Fig.~\ref{fig:coordinate}, the two polarizations are $h_{+}^{\rm TT} \equiv h_{xx}^{\rm TT} = -h_{yy}^{\rm TT}$ and $h_{\cross} \equiv h_{xy}^{\rm TT} =h_{yx}^{\rm TT}$. Eq.~(\ref{eq:strain_of_memory}) can be written as
		\begin{equation}
		\begin{aligned}
			\Delta h_{+}^{\rm TT}+ i \Delta h_{\cross}^{\rm TT} &= \frac{2}{d} \int_{-\infty}^{t-d} L_\nu\left(t'\right) dt' \\ 
			&\int {F\left(t', \Omega'\right)}\left( 1+ \cos{\theta}  \right)e^{i2\phi}  \: d\Omega'\: ,
		\label{eq:strain_of_memory_2}
		\end{aligned}
		\end{equation}
where $\theta$ is the angle between the flux going into $d\Omega'$ and the direction to the detector, and $\phi$ is the azimuthal angle with respect to the $x$ axis in the $x-y$ plane. From Eq.~(\ref{eq:strain_of_memory_2}) it's clear that the rise time for the non-oscillatory gravitational wave memory signal to reach its final strain is the same as the duration of the neutrino burst in the detector's rest frame.

%------------------------------------------------------------------------------------------------------------------------
% Appendix B
%------------------------------------------------------------------------------------------------------------------------

\section{Detector's Response to a GWM Signal}\label{appendix b}

\begin{figure}[t]
\includegraphics[width=1.0\linewidth]{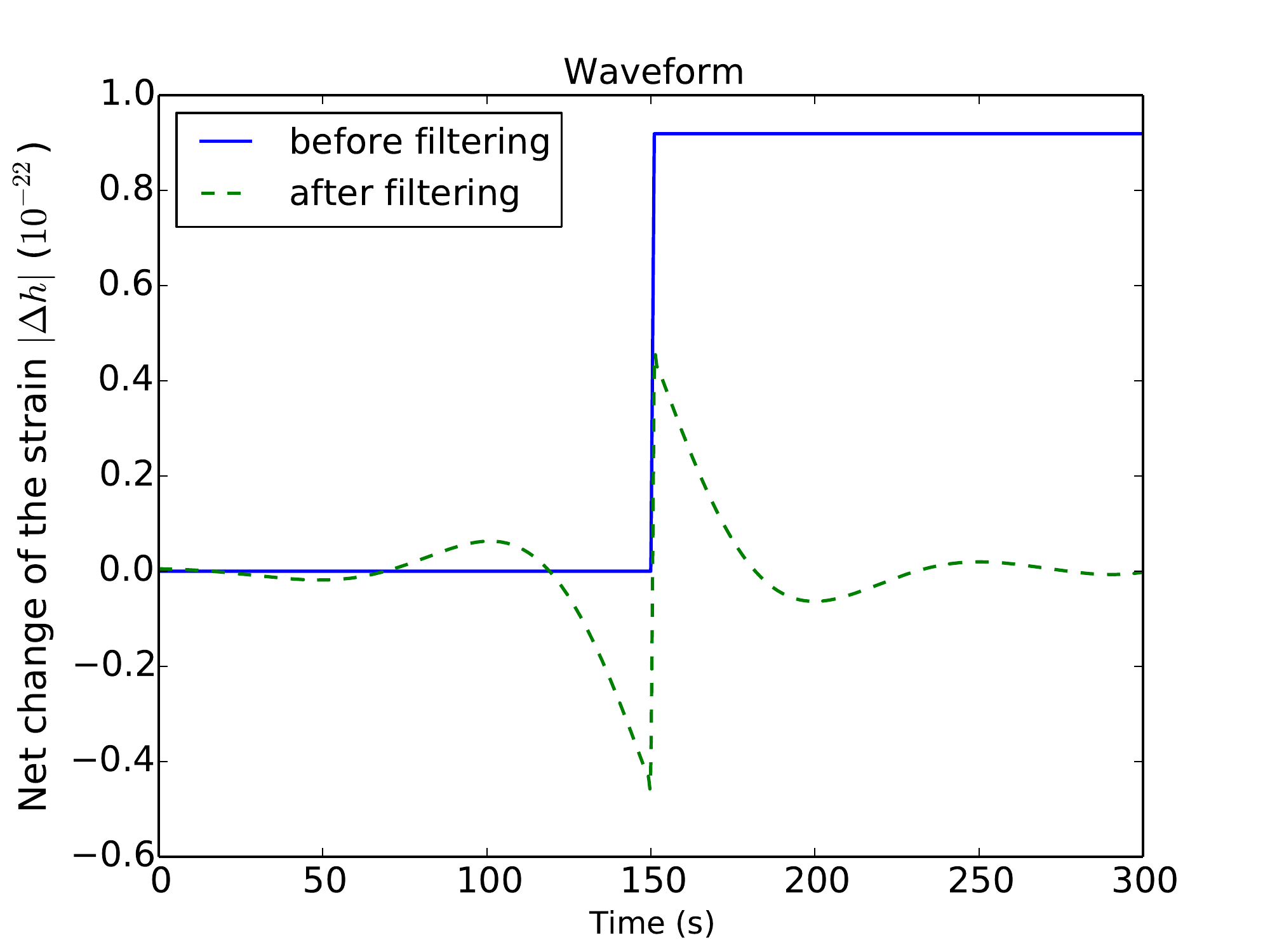}
\caption{Estimated gravitational wave time series from a SMS collapse with $M^{\rm HC} = 10^5M_\odot$ at $z=7$ and the neutrino emission asymmetry $\alpha = -0.02$. 
The detector is placed at equatorial plane $\xi = \pi/2$. 
The blue solid line shows the detector's arm response to GWM signal. 
The green dashed line shows the time series filtered with a $10^{-2}-10^0$ Hz bandpass filter to illustrate the signal seen by DECIGO.}
\label{fig:bandpass}
\end{figure}

The GWM is a non-oscillatory signal which causes a permanent displacement of the detector's arm length after this wave train has passed. Its effect on two freely falling masses is a ``DC" offset-like signal, with the rise time equal to the signal burst time $\Delta t_m$. As an example, we can estimate the strain as a function of detector frame time based on the total neutrino emission and burst time taken from the Shi \& Fuller neutrino luminosity and timescale result. This estimate results in the solid curve in Fig.~\ref{fig:bandpass}. It shows the full GWM waveform for $M^{\rm HC} = 10^5\,M_\odot$ SMS collapse at $z=7$.

Laser interferometry gravitational wave detectors' sensitivity curves are frequency dependent, so only a narrow frequency band is ``audible'' to such detectors. To mimic DECIGO's response to the GWM signal, we use a bandpass filter in the frequency band [0.01, 1]~Hz. The green dashed curve in Fig.~\ref{fig:bandpass} represents the response of DECIGO to the GWM signal (the solid curve). The waveform after the filtering will not resemble a ``DC''-like signal because of the suppression of low frequencies.

%%%%%%%%%%%%%%%%%%%%
%%---------------- Figure 6 ------------------   
%%---------------- Figure 6 ------------------   
%%%%%%%%%%%%%%%%%%%%

%------------------------------------------------------------------------------------------------------------------------
% Appendix C
%------------------------------------------------------------------------------------------------------------------------
\section{Neutrino Absorption by BH trapped Surface}\label{appendix c}

Local neutrino emissivity ($\propto T^9$) and overall neutrino luminosity both increase as the core collapses and the temperature increases. Peak neutrino luminosity will occur very near where a trapped surface forms and gravitational redshift rather abruptly cuts off neutrino radiation to infinity. Just what that peak luminosity is and, concomitantly, the amplitude of the GWM signal both depend on details of relativistic effects near black hole formation. With our nearly Newtonian treatment, we can make only cogent, order of magnitude estimates, of these effects. The essence of the problem: The competition between increasing neutrino emissivity  and gravitational redshift implies that most of the neutrinos are emitted at a thin spherical shell of radius somewhere between $1M^{\rm HC}$ and $2M^{\rm HC}$ within the dynamical time scale $2M^{\rm HC}$. The neutrino luminosity calculation from the Shi \& Fuller result is based on the assumption that neutrinos only move radially outward. Yet a significant fraction of neutrinos that move radially inward will not have enough time to pass through the Schwarzschild radius $2M^{\rm HC}$ at the onset of BH formation. The consequences are: (1) the actual neutrino luminosity is smaller than what is calculated in Shi \& Fuller and (2) only the neutrinos that are not trapped in the BH can contribute the neutrino emission asymmetry via the ISW-like effect.

%%%%%%%%%%%%%%%%%%%% 
%%---------------- Figure 7 ------------------   
\begin{figure}[t]
\includegraphics[width=0.55\linewidth]{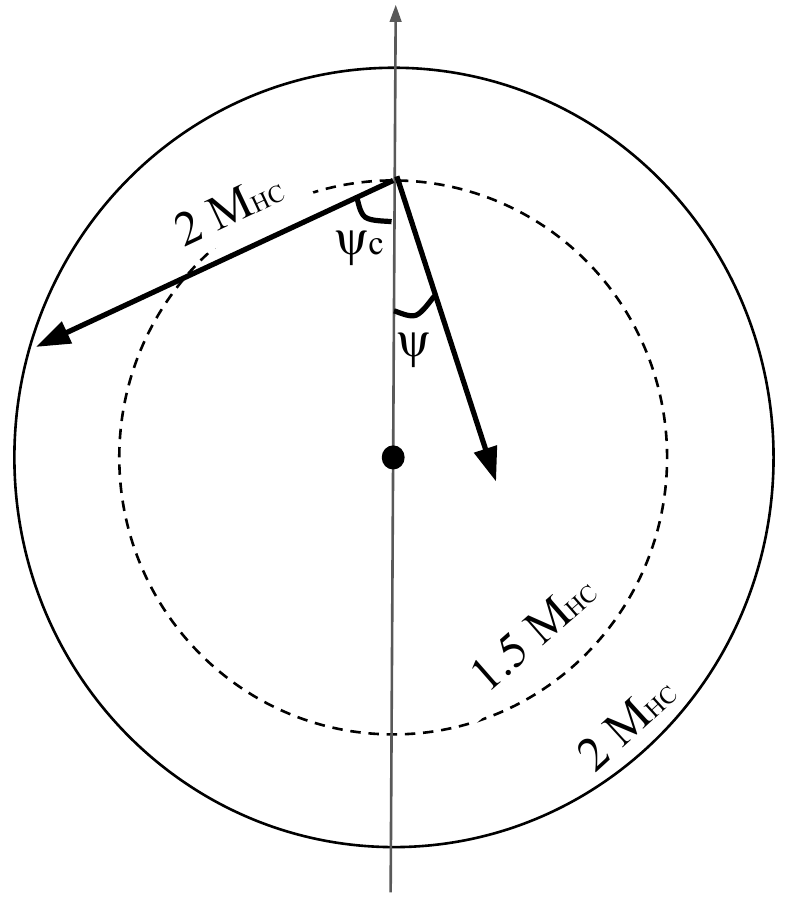}
\caption{Illustration of neutrino absorption by BH trapped surface formation. The solid circle is the trapped surface after the BH is formed. The dashed circle is the radius of neutrino peak production site. Neutrinos moving toward the core at the angle $\psi<\psi_c$ will be trapped in the BH and therefore make no contribution to the neutrino emission asymmetry $\alpha$.}
\label{fig:trapped_surface}
\end{figure}
%%---------------- Figure 7 ------------------   
%%%%%%%%%%%%%%%%%%%%

\begin{figure}[t]
\includegraphics[width=1.0\linewidth]{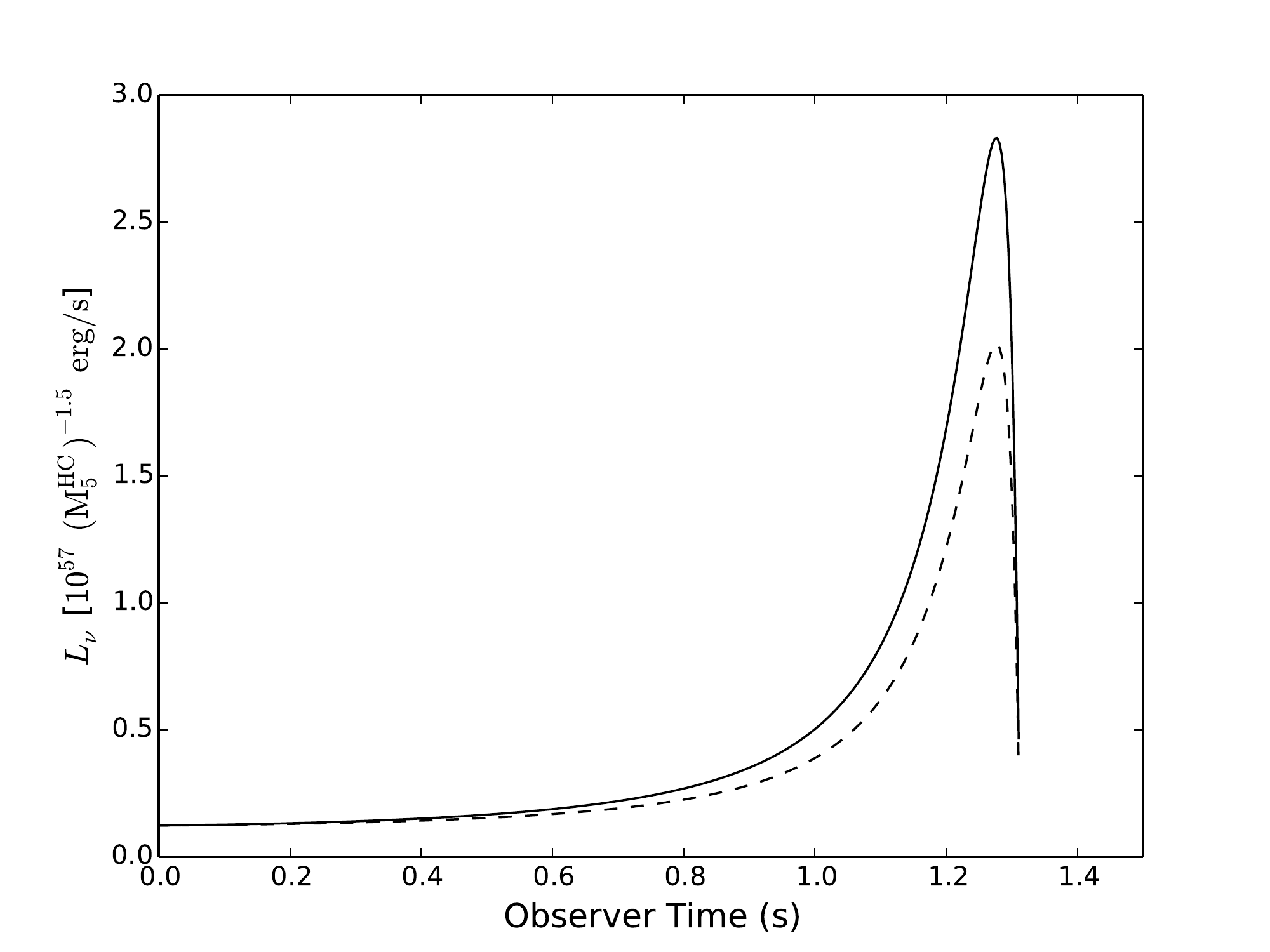}
\caption{Time evolution of the neutrino luminosity. The solid curve is taken from Shi \& Fuller, which assumes that neutrinos only move radially outward. The dashed curve takes account the neutrino luminosity loss due to the different time of flight at different emission angles.}
\label{fig:neutrino_luminosity}
\end{figure}

Assume that peak neutrino emission happens in a thin spherical shell with the radius $r_{\rm peak} \approx 1.5 M^{\rm HC}$ within one dynamical time before the BH formation. Neutrinos emitted into an inwardly-directed pencil of directions with launch angle $\psi$ (relative to radially inward-directed unit vector) smaller than the critical angle $\psi_c$ have a time of flight greater than $2M^{\rm HC}$ and therefore will be inside the trapped surface when the BH is formed. The critical angle can be estimated easily in Euclidean geometry: $\psi_c \approx 68^\circ$ (see Fig.~\ref{fig:trapped_surface}). As a result, the fraction of neutrino luminosity loss, $\epsilon$, is approximately 
	\begin{equation} 
		\epsilon =\frac{1}{4\pi} \int_0^{2\pi} \int_0^{\psi_c} \sin\theta d\theta d\phi \approx 0.3
	\end{equation}
Fig.~\ref{fig:neutrino_luminosity} shows the time evolution of neutrino luminosity for an observer at infinity. The solid curve is taken from the Shi \& Fuller result, which assumes all neutrinos move radially outward. The dashed curve is the neutrino luminosity after taking account of the energy loss $\epsilon$ stemming from different time of flight along trajectories at different emission angles.

Here we show a crude estimate simply to illustrate how the BH trapped surface could change the neutrino luminosity. Certainly, more sophisticated, fully relativistic 3-dimensional hydrodynamical simulations together with a numerical spacetime/gravitational wave calculation are warranted. Nevertheless, throughout this paper, to facilitate a parameter survey and to illustrate the basic effects we adopt the neutrino luminosity functions taken from the Shi \& Fuller and Linke et al. results.

%%%%%%%%%%%%%%%%%%%% 
%%---------------- Figure 8 ------------------   
%%---------------- Figure 8 ------------------   
%%%%%%%%%%%%%%%%%%%%

Neutrino trapping by trapped surface formation also impacts estimates of the neutrino emission asymmetry. Neutrino born on trajectories with angle $\psi$ less than the critical angle $\psi_c$ will be trapped in the BH. These will not make any contribution to the neutrino emission asymmetry $\alpha$. Neutrinos moving on trajectories with angles $\psi > \pi/2$ will not experience any ISW-like effect because they do not stream into the collapsing core. Only neutrinos moving on trajectory angles between $\psi_c$ and $\pi/2$ can experience the differential blueshift-redshift effect and still escape to contribute to the neutrino emission asymmetry.

Using the Newtonian picture, a neutrino directed toward the collapsing core of the SMS will lose a fraction of its energy, $\delta E / E \sim \delta M / r$, from the ISW-like, angle-dependent, differential redshift-blueshift effect. The timescale necessary for the neutrino to stream through the core and back to its initial radius is $\delta t = 2 r \cos \psi$, where $\psi$ is the angle between the neutrino trajectory and the radial line. The increase in the enclosed mass is $\delta M \sim \bar\rho \times (4 \pi r^2) \times (2 r \cos \psi)$, where $\bar\rho$ is the average density of the homologous core close to the BH formation. The fractional energy loss is $\delta E/E \sim \left({r}/{r_s}\right)\cos \psi$.

Most of the neutrinos are emitted in a relatively thin spherical shell of radius somewhere at $r_{\rm peak} \approx 1.5 M^{\rm HC}$. Consequently, in this paper, we approximate the fractional energy loss in a radius-independent form $\delta E/E \sim {3\over4}\times\cos \psi$. Note that this function is only meant to represent the energy loss due to the ISW-like effect. It does not include the absorption accompanying trapped surface formation.

As a simple model, let the neutrino emissivity in the peak emission shell be parameterized by $Q_\eta \left(\theta\right) = Q_0 \left(1 + \eta \cos^2 \theta\right)$, where $\eta$ is the neutrino emissivity asymmetry between volume elements along the polar direction and the equatorial plane and $Q_0$ is proportional to the volume-averaged emissivity, $\langle Q_\eta \rangle = Q_0 \left(1 + \eta /3 \right)$. To estimate the polar-equatorial neutrino energy flux anisotropy, we need to estimate the neutrino energy fluxes that experience the differential blueshift-redshift effect and stream into a solid angle $d \Omega$ in the polar direction, along the negative z-axis ($\phi^{\rm (pol)}$), and an equatorial direction, along the negative y-axis ($\phi^{\rm (eq)}$):
%%%%%%%%%%%%%
%    Equation 1 & 2    %%
%%%%%%%%%%%%%
		\begin{equation}
			\phi^{\rm (pol)} =  \left. \int\limits_0^{2\pi} \int\limits_{\psi_c}^{\pi\over2} \left(1-{3\over4}\cos \theta \right) \,Q_\eta \left(\theta\right) \sin \theta \,d \theta \, d\phi \right. ,
			\label{eq:polar}
		\end{equation}
		\begin{equation}
			\phi^{\rm (eq)} =  \left. \int\limits_0^{2\pi} \int\limits_{\psi_c}^{\pi\over2} \left(1- {3\over4}\cos \theta' \right) \,Q_\eta \left(\theta\right) \sin \theta' \,d \theta' \, d\phi'  \right. ,
			\label{eq:equator}
		\end{equation}
where $\theta$ is the polar angle from z-direction and $\theta'$ is the new polar angle measured from y-direction. Here we take $\psi_c$ to be $68^\circ$. Parameterizing the polar-equatorial neutrino energy flux asymmetry, $\alpha$, as the ratio of the total flux in the polar direction to the flux in the equatorial directions subtracted by unity, and using $\eta = 0.25$, we estimate a neutrino emission asymmetry:
		\begin{equation}
			\alpha = \frac{2\pi \langle Q_\eta\rangle + \phi^{\rm (pol)}}{2\pi \langle Q_\eta\rangle+ \phi^{\rm (eq)}} -1 \approx  -0.02.
			\label{eq:nu_asym}
		\end{equation}

The emission asymmetry parameter $\alpha$ will depend on the SMS initial rotation speed. A faster rotation will induce a larger emissivity asymmetry, leading to a larger neutrino emission asymmetry. But given that the initial state of the SMS is unknown, $\alpha$ should be treated as a free parameter. Nevertheless, for illustrative purposes we will use $\alpha = - 0.02$ throughout this paper.

%------------------------------------------------------------------------------------------------------------------------
% Bibliography
%------------------------------------------------------------------------------------------------------------------------
\bibliography{GWM_SMS}

\begin{thebibliography}{68}%
\makeatletter
\providecommand \@ifxundefined [1]{%
 \@ifx{#1\undefined}
}%
\providecommand \@ifnum [1]{%
 \ifnum #1\expandafter \@firstoftwo
 \else \expandafter \@secondoftwo
 \fi
}%
\providecommand \@ifx [1]{%
 \ifx #1\expandafter \@firstoftwo
 \else \expandafter \@secondoftwo
 \fi
}%
\providecommand \natexlab [1]{#1}%
\providecommand \enquote  [1]{``#1''}%
\providecommand \bibnamefont  [1]{#1}%
\providecommand \bibfnamefont [1]{#1}%
\providecommand \citenamefont [1]{#1}%
\providecommand \href@noop [0]{\@secondoftwo}%
\providecommand \href [0]{\begingroup \@sanitize@url \@href}%
\providecommand \@href[1]{\@@startlink{#1}\@@href}%
\providecommand \@@href[1]{\endgroup#1\@@endlink}%
\providecommand \@sanitize@url [0]{\catcode `\\12\catcode `\$12\catcode
  `\&12\catcode `\#12\catcode `\^12\catcode `\_12\catcode `\%12\relax}%
\providecommand \@@startlink[1]{}%
\providecommand \@@endlink[0]{}%
\providecommand \url  [0]{\begingroup\@sanitize@url \@url }%
\providecommand \@url [1]{\endgroup\@href {#1}{\urlprefix }}%
\providecommand \urlprefix  [0]{URL }%
\providecommand \Eprint [0]{\href }%
\providecommand \doibase [0]{http://dx.doi.org/}%
\providecommand \selectlanguage [0]{\@gobble}%
\providecommand \bibinfo  [0]{\@secondoftwo}%
\providecommand \bibfield  [0]{\@secondoftwo}%
\providecommand \translation [1]{[#1]}%
\providecommand \BibitemOpen [0]{}%
\providecommand \bibitemStop [0]{}%
\providecommand \bibitemNoStop [0]{.\EOS\space}%
\providecommand \EOS [0]{\spacefactor3000\relax}%
\providecommand \BibitemShut  [1]{\csname bibitem#1\endcsname}%
\let\auto@bib@innerbib\@empty
%</preamble>
\bibitem [{\citenamefont {Seto}\ \emph {et~al.}(2001)\citenamefont {Seto},
  \citenamefont {Kawamura},\ and\ \citenamefont {Nakamura}}]{Seto:2001qf}%
  \BibitemOpen
  \bibfield  {author} {\bibinfo {author} {\bibfnamefont {N.}~\bibnamefont
  {Seto}}, \bibinfo {author} {\bibfnamefont {S.}~\bibnamefont {Kawamura}}, \
  and\ \bibinfo {author} {\bibfnamefont {T.}~\bibnamefont {Nakamura}},\ }\href
  {\doibase 10.1103/PhysRevLett.87.221103} {\bibfield  {journal} {\bibinfo
  {journal} {Phys. Rev. Lett.}\ }\textbf {\bibinfo {volume} {87}},\ \bibinfo
  {pages} {221103} (\bibinfo {year} {2001})},\ \Eprint
  {http://arxiv.org/abs/astro-ph/0108011} {arXiv:astro-ph/0108011 [astro-ph]}
  \BibitemShut {NoStop}%
%%CITATION = ASTRO-PH/0108011;%%
\bibitem [{\citenamefont {Kawamura}\ \emph {et~al.}(2011)\citenamefont
  {Kawamura} \emph {et~al.}}]{Kawamura:2011zz}%
  \BibitemOpen
  \bibfield  {author} {\bibinfo {author} {\bibfnamefont {S.}~\bibnamefont
  {Kawamura}} \emph {et~al.},\ }\href {\doibase 10.1088/0264-9381/28/9/094011}
  {\bibfield  {journal} {\bibinfo  {journal} {Class. Quant. Grav.}\ }\textbf
  {\bibinfo {volume} {28}},\ \bibinfo {pages} {094011} (\bibinfo {year}
  {2011})}\BibitemShut {NoStop}%
%%CITATION = CQGRD,28,094011;%%
\bibitem [{\citenamefont {Hoyle}\ and\ \citenamefont
  {Fowler}(1963)}]{Foyle:1963aa}%
  \BibitemOpen
  \bibfield  {author} {\bibinfo {author} {\bibfnamefont {F.}~\bibnamefont
  {Hoyle}}\ and\ \bibinfo {author} {\bibfnamefont {W.~S.}\ \bibnamefont
  {Fowler}},\ }\href {\doibase doi:10.1038/197533a0} {\bibfield  {journal}
  {\bibinfo  {journal} {Nature}\ }\textbf {\bibinfo {volume} {197}},\ \bibinfo
  {pages} {533} (\bibinfo {year} {1963})}\BibitemShut {NoStop}%
\bibitem [{\citenamefont {Iben}(1963)}]{Iben:1963}%
  \BibitemOpen
  \bibfield  {author} {\bibinfo {author} {\bibfnamefont {I.}~\bibnamefont
  {Iben}},\ }\href {\doibase 10.1086/147708} {\bibfield  {journal} {\bibinfo
  {journal} {Astrophys. J.}\ }\textbf {\bibinfo {volume} {138}},\ \bibinfo
  {pages} {1090} (\bibinfo {year} {1963})}\BibitemShut {NoStop}%
\bibitem [{\citenamefont {Fowler}(1964)}]{Fowler:1964zza}%
  \BibitemOpen
  \bibfield  {author} {\bibinfo {author} {\bibfnamefont {W.~A.}\ \bibnamefont
  {Fowler}},\ }\href {\doibase 10.1103/RevModPhys.36.545} {\bibfield  {journal}
  {\bibinfo  {journal} {Rev. Mod. Phys.}\ }\textbf {\bibinfo {volume} {36}},\
  \bibinfo {pages} {545} (\bibinfo {year} {1964})}\BibitemShut {NoStop}%
%%CITATION = RMPHA,36,545;%%
\bibitem [{\citenamefont {Zeldovich}(1970)}]{Zeldovich:1969sb}%
  \BibitemOpen
  \bibfield  {author} {\bibinfo {author} {\bibfnamefont {{\relax Ya}.~B.}\
  \bibnamefont {Zeldovich}},\ }\href@noop {} {\bibfield  {journal} {\bibinfo
  {journal} {Astron. Astrophys.}\ }\textbf {\bibinfo {volume} {5}},\ \bibinfo
  {pages} {84} (\bibinfo {year} {1970})}\BibitemShut {NoStop}%
%%CITATION = AAEJA,5,84;%%
\bibitem [{\citenamefont {Fuller}\ \emph {et~al.}(1986)\citenamefont {Fuller},
  \citenamefont {Woosley},\ and\ \citenamefont {Weaver}}]{Fuller:1986}%
  \BibitemOpen
  \bibfield  {author} {\bibinfo {author} {\bibfnamefont {G.~M.}\ \bibnamefont
  {Fuller}}, \bibinfo {author} {\bibfnamefont {S.~E.}\ \bibnamefont {Woosley}},
  \ and\ \bibinfo {author} {\bibfnamefont {T.~A.}\ \bibnamefont {Weaver}},\
  }\href {\doibase 10.1086/164452} {\bibfield  {journal} {\bibinfo  {journal}
  {Astrophys. J.}\ }\textbf {\bibinfo {volume} {307}},\ \bibinfo {pages} {675}
  (\bibinfo {year} {1986})}\BibitemShut {NoStop}%
\bibitem [{\citenamefont {Chandrasekhar}(1964)}]{Chandrasekhar:1964zz}%
  \BibitemOpen
  \bibfield  {author} {\bibinfo {author} {\bibfnamefont {S.}~\bibnamefont
  {Chandrasekhar}},\ }\href {\doibase 10.1086/147938} {\bibfield  {journal}
  {\bibinfo  {journal} {Astrophys. J.}\ }\textbf {\bibinfo {volume} {140}},\
  \bibinfo {pages} {417} (\bibinfo {year} {1964})},\ \bibinfo {note} {[Erratum:
  Astrophys. J. 140,1342 (1964)]}\BibitemShut {NoStop}%
%%CITATION = ASJOA,140,417;%%
\bibitem [{\citenamefont {Feynman}(1996)}]{Feynman:1996kb}%
  \BibitemOpen
  \bibfield  {author} {\bibinfo {author} {\bibfnamefont {R.~P.}\ \bibnamefont
  {Feynman}},\ }\href@noop {} {\emph {\bibinfo {title} {{Feynman lectures on
  gravitation}}}},\ edited by\ \bibinfo {editor} {\bibfnamefont {F.~B.}\
  \bibnamefont {Morinigo}}\ and\ \bibinfo {editor} {\bibfnamefont {W.~G.}\
  \bibnamefont {Wagner}}\ (\bibinfo {year} {1996})\BibitemShut {NoStop}%
%%CITATION = INSPIRE-427379;%%
\bibitem [{\citenamefont {Mortlock}\ \emph {et~al.}(2011)\citenamefont
  {Mortlock} \emph {et~al.}}]{Mortlock:2011va}%
  \BibitemOpen
  \bibfield  {author} {\bibinfo {author} {\bibfnamefont {D.~J.}\ \bibnamefont
  {Mortlock}} \emph {et~al.},\ }\href {\doibase 10.1038/nature10159} {\bibfield
   {journal} {\bibinfo  {journal} {Nature}\ }\textbf {\bibinfo {volume}
  {474}},\ \bibinfo {pages} {616} (\bibinfo {year} {2011})},\ \Eprint
  {http://arxiv.org/abs/1106.6088} {arXiv:1106.6088 [astro-ph.CO]} \BibitemShut
  {NoStop}%
%%CITATION = ARXIV:1106.6088;%%
\bibitem [{\citenamefont {Venemans}\ \emph {et~al.}(2013)\citenamefont
  {Venemans} \emph {et~al.}}]{Venemans:2013npa}%
  \BibitemOpen
  \bibfield  {author} {\bibinfo {author} {\bibfnamefont {B.~P.}\ \bibnamefont
  {Venemans}} \emph {et~al.},\ }\href {\doibase 10.1088/0004-637X/779/1/24}
  {\bibfield  {journal} {\bibinfo  {journal} {Astrophys. J.}\ }\textbf
  {\bibinfo {volume} {779}},\ \bibinfo {pages} {24} (\bibinfo {year} {2013})},\
  \Eprint {http://arxiv.org/abs/1311.3666} {arXiv:1311.3666 [astro-ph.CO]}
  \BibitemShut {NoStop}%
%%CITATION = ARXIV:1311.3666;%%
\bibitem [{\citenamefont {Wu}\ \emph {et~al.}(2015)\citenamefont {Wu} \emph
  {et~al.}}]{Wu:2015}%
  \BibitemOpen
  \bibfield  {author} {\bibinfo {author} {\bibfnamefont {X.-B.}\ \bibnamefont
  {Wu}} \emph {et~al.},\ }\href {\doibase 10.1038/nature14241} {\bibfield
  {journal} {\bibinfo  {journal} {Nature}\ }\textbf {\bibinfo {volume} {518}},\
  \bibinfo {pages} {512} (\bibinfo {year} {2015})},\ \Eprint
  {http://arxiv.org/abs/1502.07418} {arXiv:1502.07418 [astro-GA]} \BibitemShut
  {NoStop}%
%%CITATION = ARXIV:1502.07418;%%
\bibitem [{\citenamefont {Ba{\~n}ados}\ \emph {et~al.}(2017)\citenamefont
  {Ba{\~n}ados} \emph {et~al.}}]{Banados:2017}%
  \BibitemOpen
  \bibfield  {author} {\bibinfo {author} {\bibfnamefont {E.}~\bibnamefont
  {Ba{\~n}ados}} \emph {et~al.},\ }\href {\doibase 10.1038/nature25180}
  {\bibfield  {journal} {\bibinfo  {journal} {Nature}\ } (\bibinfo {year}
  {2017}),\ 10.1038/nature25180}\BibitemShut {NoStop}%
\bibitem [{\citenamefont {{Venemans}}\ \emph {et~al.}(2017)\citenamefont
  {{Venemans}} \emph {et~al.}}]{2017arXiv171201886V}%
  \BibitemOpen
  \bibfield  {author} {\bibinfo {author} {\bibfnamefont {B.}~\bibnamefont
  {{Venemans}}} \emph {et~al.},\ }\href {\doibase 10.3847/2041-8213/aa943a}
  {\bibfield  {journal} {\bibinfo  {journal} {Astrophys. J.}\ }\textbf
  {\bibinfo {volume} {851}},\ \bibinfo {pages} {L8} (\bibinfo {year} {2017})},\
  \Eprint {http://arxiv.org/abs/1712.01886} {arXiv:1712.01886 [astro-ph.GA]}
  \BibitemShut {NoStop}%
\bibitem [{\citenamefont {Begelman}\ and\ \citenamefont
  {Rees}(1978)}]{Begelman:1978}%
  \BibitemOpen
  \bibfield  {author} {\bibinfo {author} {\bibfnamefont {M.~C.}\ \bibnamefont
  {Begelman}}\ and\ \bibinfo {author} {\bibfnamefont {M.~J.}\ \bibnamefont
  {Rees}},\ }\href {\doibase 10.1093/mnras/185.4.847} {\bibfield  {journal}
  {\bibinfo  {journal} {Mon. Not. Roy. Astron. Soc.}\ }\textbf {\bibinfo
  {volume} {185}},\ \bibinfo {pages} {847} (\bibinfo {year}
  {1978})}\BibitemShut {NoStop}%
\bibitem [{\citenamefont {Rees}(1984)}]{Rees:1984si}%
  \BibitemOpen
  \bibfield  {author} {\bibinfo {author} {\bibfnamefont {M.~J.}\ \bibnamefont
  {Rees}},\ }\href {\doibase 10.1146/annurev.aa.22.090184.002351} {\bibfield
  {journal} {\bibinfo  {journal} {Ann. Rev. Astron. Astrophys.}\ }\textbf
  {\bibinfo {volume} {22}},\ \bibinfo {pages} {471} (\bibinfo {year}
  {1984})}\BibitemShut {NoStop}%
%%CITATION = ARAAA,22,471;%%
\bibitem [{\citenamefont {Haehnelt}\ and\ \citenamefont
  {Rees}(1993)}]{Haehnelt:1993yy}%
  \BibitemOpen
  \bibfield  {author} {\bibinfo {author} {\bibfnamefont {M.~G.}\ \bibnamefont
  {Haehnelt}}\ and\ \bibinfo {author} {\bibfnamefont {M.~J.}\ \bibnamefont
  {Rees}},\ }\href@noop {} {\bibfield  {journal} {\bibinfo  {journal} {Mon.
  Not. Roy. Astron. Soc.}\ }\textbf {\bibinfo {volume} {263}},\ \bibinfo
  {pages} {168} (\bibinfo {year} {1993})}\BibitemShut {NoStop}%
%%CITATION = MNRAA,263,168;%%
\bibitem [{\citenamefont {Loeb}\ and\ \citenamefont
  {Rasio}(1994)}]{Loeb:1994wv}%
  \BibitemOpen
  \bibfield  {author} {\bibinfo {author} {\bibfnamefont {A.}~\bibnamefont
  {Loeb}}\ and\ \bibinfo {author} {\bibfnamefont {F.~A.}\ \bibnamefont
  {Rasio}},\ }\href {\doibase 10.1086/174548} {\bibfield  {journal} {\bibinfo
  {journal} {Astrophys. J.}\ }\textbf {\bibinfo {volume} {432}},\ \bibinfo
  {pages} {52} (\bibinfo {year} {1994})},\ \Eprint
  {http://arxiv.org/abs/astro-ph/9401026} {arXiv:astro-ph/9401026 [astro-ph]}
  \BibitemShut {NoStop}%
%%CITATION = ASTRO-PH/9401026;%%
\bibitem [{\citenamefont {Eisenstein}\ and\ \citenamefont
  {Loeb}(1995)}]{Eisenstein:1994nh}%
  \BibitemOpen
  \bibfield  {author} {\bibinfo {author} {\bibfnamefont {D.~J.}\ \bibnamefont
  {Eisenstein}}\ and\ \bibinfo {author} {\bibfnamefont {A.}~\bibnamefont
  {Loeb}},\ }\href {\doibase 10.1086/175498} {\bibfield  {journal} {\bibinfo
  {journal} {Astrophys. J.}\ }\textbf {\bibinfo {volume} {443}},\ \bibinfo
  {pages} {11} (\bibinfo {year} {1995})},\ \Eprint
  {http://arxiv.org/abs/astro-ph/9401016} {arXiv:astro-ph/9401016 [astro-ph]}
  \BibitemShut {NoStop}%
%%CITATION = ASTRO-PH/9401016;%%
\bibitem [{\citenamefont {Oh}\ and\ \citenamefont {Haiman}(2002)}]{Oh:2001ex}%
  \BibitemOpen
  \bibfield  {author} {\bibinfo {author} {\bibfnamefont {S.~P.}\ \bibnamefont
  {Oh}}\ and\ \bibinfo {author} {\bibfnamefont {Z.}~\bibnamefont {Haiman}},\
  }\href {\doibase 10.1086/339393} {\bibfield  {journal} {\bibinfo  {journal}
  {Astrophys. J.}\ }\textbf {\bibinfo {volume} {569}},\ \bibinfo {pages} {558}
  (\bibinfo {year} {2002})},\ \Eprint {http://arxiv.org/abs/astro-ph/0108071}
  {arXiv:astro-ph/0108071 [astro-ph]} \BibitemShut {NoStop}%
%%CITATION = ASTRO-PH/0108071;%%
\bibitem [{\citenamefont {Bromm}\ and\ \citenamefont
  {Loeb}(2003)}]{Bromm:2002hb}%
  \BibitemOpen
  \bibfield  {author} {\bibinfo {author} {\bibfnamefont {V.}~\bibnamefont
  {Bromm}}\ and\ \bibinfo {author} {\bibfnamefont {A.}~\bibnamefont {Loeb}},\
  }\href {\doibase 10.1086/377529} {\bibfield  {journal} {\bibinfo  {journal}
  {Astrophys. J.}\ }\textbf {\bibinfo {volume} {596}},\ \bibinfo {pages} {34}
  (\bibinfo {year} {2003})},\ \Eprint {http://arxiv.org/abs/astro-ph/0212400}
  {arXiv:astro-ph/0212400 [astro-ph]} \BibitemShut {NoStop}%
%%CITATION = ASTRO-PH/0212400;%%
\bibitem [{\citenamefont {Koushiappas}\ \emph {et~al.}(2004)\citenamefont
  {Koushiappas}, \citenamefont {Bullock},\ and\ \citenamefont
  {Dekel}}]{Koushiappas:2003zn}%
  \BibitemOpen
  \bibfield  {author} {\bibinfo {author} {\bibfnamefont {S.~M.}\ \bibnamefont
  {Koushiappas}}, \bibinfo {author} {\bibfnamefont {J.~S.}\ \bibnamefont
  {Bullock}}, \ and\ \bibinfo {author} {\bibfnamefont {A.}~\bibnamefont
  {Dekel}},\ }\href {\doibase 10.1111/j.1365-2966.2004.08190.x} {\bibfield
  {journal} {\bibinfo  {journal} {Mon. Not. Roy. Astron. Soc.}\ }\textbf
  {\bibinfo {volume} {354}},\ \bibinfo {pages} {292} (\bibinfo {year}
  {2004})},\ \Eprint {http://arxiv.org/abs/astro-ph/0311487}
  {arXiv:astro-ph/0311487 [astro-ph]} \BibitemShut {NoStop}%
%%CITATION = ASTRO-PH/0311487;%%
\bibitem [{\citenamefont {Lodato}\ and\ \citenamefont
  {Natarajan}(2006)}]{Lodato:2006hw}%
  \BibitemOpen
  \bibfield  {author} {\bibinfo {author} {\bibfnamefont {G.}~\bibnamefont
  {Lodato}}\ and\ \bibinfo {author} {\bibfnamefont {P.}~\bibnamefont
  {Natarajan}},\ }\href {\doibase 10.1111/j.1365-2966.2006.10801.x} {\bibfield
  {journal} {\bibinfo  {journal} {Mon. Not. Roy. Astron. Soc.}\ }\textbf
  {\bibinfo {volume} {371}},\ \bibinfo {pages} {1813} (\bibinfo {year}
  {2006})},\ \Eprint {http://arxiv.org/abs/astro-ph/0606159}
  {arXiv:astro-ph/0606159 [astro-ph]} \BibitemShut {NoStop}%
%%CITATION = ASTRO-PH/0606159;%%
\bibitem [{\citenamefont {Shapiro}\ and\ \citenamefont
  {Teukolsky}(1985)}]{shapiro1985collapse}%
  \BibitemOpen
  \bibfield  {author} {\bibinfo {author} {\bibfnamefont {S.~L.}\ \bibnamefont
  {Shapiro}}\ and\ \bibinfo {author} {\bibfnamefont {S.~A.}\ \bibnamefont
  {Teukolsky}},\ }\href@noop {} {\bibfield  {journal} {\bibinfo  {journal}
  {Astrophys. J.}\ }\textbf {\bibinfo {volume} {292}},\ \bibinfo {pages} {L41}
  (\bibinfo {year} {1985})}\BibitemShut {NoStop}%
\bibitem [{\citenamefont {Quinlan}\ and\ \citenamefont
  {Shapiro}(1987)}]{quinlan1987collapse}%
  \BibitemOpen
  \bibfield  {author} {\bibinfo {author} {\bibfnamefont {G.~D.}\ \bibnamefont
  {Quinlan}}\ and\ \bibinfo {author} {\bibfnamefont {S.~L.}\ \bibnamefont
  {Shapiro}},\ }\href@noop {} {\bibfield  {journal} {\bibinfo  {journal}
  {Astrophys. J.}\ }\textbf {\bibinfo {volume} {321}},\ \bibinfo {pages} {199}
  (\bibinfo {year} {1987})}\BibitemShut {NoStop}%
\bibitem [{\citenamefont {Quinlan}\ and\ \citenamefont
  {Shapiro}(1989)}]{quinlan1989dynamical}%
  \BibitemOpen
  \bibfield  {author} {\bibinfo {author} {\bibfnamefont {G.~D.}\ \bibnamefont
  {Quinlan}}\ and\ \bibinfo {author} {\bibfnamefont {S.~L.}\ \bibnamefont
  {Shapiro}},\ }\href@noop {} {\bibfield  {journal} {\bibinfo  {journal}
  {Astrophys. J.}\ }\textbf {\bibinfo {volume} {343}},\ \bibinfo {pages} {725}
  (\bibinfo {year} {1989})}\BibitemShut {NoStop}%
\bibitem [{\citenamefont {Portegies~Zwart}\ \emph {et~al.}(2004)\citenamefont
  {Portegies~Zwart}, \citenamefont {Baumgardt}, \citenamefont {Hut},
  \citenamefont {Makino},\ and\ \citenamefont
  {McMillan}}]{PortegiesZwart:2004ggg}%
  \BibitemOpen
  \bibfield  {author} {\bibinfo {author} {\bibfnamefont {S.~F.}\ \bibnamefont
  {Portegies~Zwart}}, \bibinfo {author} {\bibfnamefont {H.}~\bibnamefont
  {Baumgardt}}, \bibinfo {author} {\bibfnamefont {P.}~\bibnamefont {Hut}},
  \bibinfo {author} {\bibfnamefont {J.}~\bibnamefont {Makino}}, \ and\ \bibinfo
  {author} {\bibfnamefont {S.~L.~W.}\ \bibnamefont {McMillan}},\ }\href
  {\doibase 10.1038/nature02448} {\bibfield  {journal} {\bibinfo  {journal}
  {Nature}\ }\textbf {\bibinfo {volume} {428}},\ \bibinfo {pages} {724}
  (\bibinfo {year} {2004})},\ \Eprint {http://arxiv.org/abs/astro-ph/0402622}
  {arXiv:astro-ph/0402622 [astro-ph]} \BibitemShut {NoStop}%
%%CITATION = ASTRO-PH/0402622;%%
\bibitem [{\citenamefont {Begelman}\ \emph {et~al.}(2006)\citenamefont
  {Begelman}, \citenamefont {Volonteri},\ and\ \citenamefont
  {Rees}}]{Begelman:2006db}%
  \BibitemOpen
  \bibfield  {author} {\bibinfo {author} {\bibfnamefont {M.~C.}\ \bibnamefont
  {Begelman}}, \bibinfo {author} {\bibfnamefont {M.}~\bibnamefont {Volonteri}},
  \ and\ \bibinfo {author} {\bibfnamefont {M.~J.}\ \bibnamefont {Rees}},\
  }\href {\doibase 10.1111/j.1365-2966.2006.10467.x} {\bibfield  {journal}
  {\bibinfo  {journal} {Mon. Not. Roy. Astron. Soc.}\ }\textbf {\bibinfo
  {volume} {370}},\ \bibinfo {pages} {289} (\bibinfo {year} {2006})},\ \Eprint
  {http://arxiv.org/abs/astro-ph/0602363} {arXiv:astro-ph/0602363 [astro-ph]}
  \BibitemShut {NoStop}%
%%CITATION = ASTRO-PH/0602363;%%
\bibitem [{\citenamefont {Begelman}(2010)}]{Begelman:2010ac}%
  \BibitemOpen
  \bibfield  {author} {\bibinfo {author} {\bibfnamefont {M.~C.}\ \bibnamefont
  {Begelman}},\ }\href {\doibase 10.1111/j.1365-2966.2009.15916.x} {\bibfield
  {journal} {\bibinfo  {journal} {Mon. Not. Roy. Astrono. Soc.}\ }\textbf
  {\bibinfo {volume} {402}},\ \bibinfo {pages} {673} (\bibinfo {year}
  {2010})},\ \Eprint {http://arxiv.org/abs/0910.4398} {arXiv:0910.4398
  [astro-ph.CO]} \BibitemShut {NoStop}%
\bibitem [{\citenamefont {Shibata}\ \emph {et~al.}(2016)\citenamefont
  {Shibata}, \citenamefont {Sekiguchi}, \citenamefont {Uchida},\ and\
  \citenamefont {Umeda}}]{Shibata:2016vzw}%
  \BibitemOpen
  \bibfield  {author} {\bibinfo {author} {\bibfnamefont {M.}~\bibnamefont
  {Shibata}}, \bibinfo {author} {\bibfnamefont {Y.}~\bibnamefont {Sekiguchi}},
  \bibinfo {author} {\bibfnamefont {H.}~\bibnamefont {Uchida}}, \ and\ \bibinfo
  {author} {\bibfnamefont {H.}~\bibnamefont {Umeda}},\ }\href {\doibase
  10.1103/PhysRevD.94.021501} {\bibfield  {journal} {\bibinfo  {journal} {Phys.
  Rev.}\ }\textbf {\bibinfo {volume} {D94}},\ \bibinfo {pages} {021501}
  (\bibinfo {year} {2016})},\ \Eprint {http://arxiv.org/abs/1606.07147}
  {arXiv:1606.07147 [astro-ph.HE]} \BibitemShut {NoStop}%
%%CITATION = ARXIV:1606.07147;%%
\bibitem [{\citenamefont {Burrows}\ and\ \citenamefont
  {Hayes}(1996)}]{Burrows:1995bb}%
  \BibitemOpen
  \bibfield  {author} {\bibinfo {author} {\bibfnamefont {A.}~\bibnamefont
  {Burrows}}\ and\ \bibinfo {author} {\bibfnamefont {J.}~\bibnamefont
  {Hayes}},\ }\href {\doibase 10.1103/PhysRevLett.76.352} {\bibfield  {journal}
  {\bibinfo  {journal} {Phys. Rev. Lett.}\ }\textbf {\bibinfo {volume} {76}},\
  \bibinfo {pages} {352} (\bibinfo {year} {1996})},\ \Eprint
  {http://arxiv.org/abs/astro-ph/9511106} {arXiv:astro-ph/9511106 [astro-ph]}
  \BibitemShut {NoStop}%
%%CITATION = ASTRO-PH/9511106;%%
\bibitem [{\citenamefont {Mueller}\ and\ \citenamefont
  {Janka}(1997)}]{Mueller:1997}%
  \BibitemOpen
  \bibfield  {author} {\bibinfo {author} {\bibfnamefont {E.}~\bibnamefont
  {Mueller}}\ and\ \bibinfo {author} {\bibfnamefont {H.-T.}\ \bibnamefont
  {Janka}},\ }\href@noop {} {\bibfield  {journal} {\bibinfo  {journal} {Astron.
  Astrophys.}\ }\textbf {\bibinfo {volume} {317,}},\ \bibinfo {pages} {140}
  (\bibinfo {year} {1997})}\BibitemShut {NoStop}%
\bibitem [{\citenamefont {Fryer}\ \emph {et~al.}(2004)\citenamefont {Fryer},
  \citenamefont {Holz},\ and\ \citenamefont {Hughes}}]{Fryer:2004wi}%
  \BibitemOpen
  \bibfield  {author} {\bibinfo {author} {\bibfnamefont {C.~L.}\ \bibnamefont
  {Fryer}}, \bibinfo {author} {\bibfnamefont {D.~E.}\ \bibnamefont {Holz}}, \
  and\ \bibinfo {author} {\bibfnamefont {S.~A.}\ \bibnamefont {Hughes}},\
  }\href {\doibase 10.1086/421040} {\bibfield  {journal} {\bibinfo  {journal}
  {Astrophys. J.}\ }\textbf {\bibinfo {volume} {609}},\ \bibinfo {pages} {288}
  (\bibinfo {year} {2004})},\ \Eprint {http://arxiv.org/abs/astro-ph/0403188}
  {arXiv:astro-ph/0403188 [astro-ph]} \BibitemShut {NoStop}%
%%CITATION = ASTRO-PH/0403188;%%
\bibitem [{\citenamefont {Ott}\ \emph {et~al.}(2006)\citenamefont {Ott},
  \citenamefont {Burrows}, \citenamefont {Dessart},\ and\ \citenamefont
  {Livne}}]{Ott:2006qp}%
  \BibitemOpen
  \bibfield  {author} {\bibinfo {author} {\bibfnamefont {C.~D.}\ \bibnamefont
  {Ott}}, \bibinfo {author} {\bibfnamefont {A.}~\bibnamefont {Burrows}},
  \bibinfo {author} {\bibfnamefont {L.}~\bibnamefont {Dessart}}, \ and\
  \bibinfo {author} {\bibfnamefont {E.}~\bibnamefont {Livne}},\ }\href
  {\doibase 10.1103/PhysRevLett.96.201102} {\bibfield  {journal} {\bibinfo
  {journal} {Phys. Rev. Lett.}\ }\textbf {\bibinfo {volume} {96}},\ \bibinfo
  {pages} {201102} (\bibinfo {year} {2006})},\ \Eprint
  {http://arxiv.org/abs/astro-ph/0605493} {arXiv:astro-ph/0605493 [astro-ph]}
  \BibitemShut {NoStop}%
%%CITATION = ASTRO-PH/0605493;%%
\bibitem [{\citenamefont {Dessart}\ \emph {et~al.}(2006)\citenamefont
  {Dessart}, \citenamefont {Burrows}, \citenamefont {Ott}, \citenamefont
  {Livne}, \citenamefont {Yoon},\ and\ \citenamefont
  {Langer}}]{Dessart:2006gd}%
  \BibitemOpen
  \bibfield  {author} {\bibinfo {author} {\bibfnamefont {L.}~\bibnamefont
  {Dessart}}, \bibinfo {author} {\bibfnamefont {A.}~\bibnamefont {Burrows}},
  \bibinfo {author} {\bibfnamefont {C.}~\bibnamefont {Ott}}, \bibinfo {author}
  {\bibfnamefont {E.}~\bibnamefont {Livne}}, \bibinfo {author} {\bibfnamefont
  {S.-C.}\ \bibnamefont {Yoon}}, \ and\ \bibinfo {author} {\bibfnamefont
  {N.}~\bibnamefont {Langer}},\ }\href {\doibase 10.1086/503626} {\bibfield
  {journal} {\bibinfo  {journal} {Astrophys. J.}\ }\textbf {\bibinfo {volume}
  {644}},\ \bibinfo {pages} {1063} (\bibinfo {year} {2006})},\ \Eprint
  {http://arxiv.org/abs/astro-ph/0601603} {arXiv:astro-ph/0601603 [astro-ph]}
  \BibitemShut {NoStop}%
%%CITATION = ASTRO-PH/0601603;%%
\bibitem [{\citenamefont {Kotake}\ \emph {et~al.}(2007)\citenamefont {Kotake},
  \citenamefont {Ohnishi},\ and\ \citenamefont {Yamada}}]{Kotake:2006aq}%
  \BibitemOpen
  \bibfield  {author} {\bibinfo {author} {\bibfnamefont {K.}~\bibnamefont
  {Kotake}}, \bibinfo {author} {\bibfnamefont {N.}~\bibnamefont {Ohnishi}}, \
  and\ \bibinfo {author} {\bibfnamefont {S.}~\bibnamefont {Yamada}},\ }\href
  {\doibase 10.1086/509320} {\bibfield  {journal} {\bibinfo  {journal}
  {Astrophys. J.}\ }\textbf {\bibinfo {volume} {655}},\ \bibinfo {pages} {406}
  (\bibinfo {year} {2007})},\ \Eprint {http://arxiv.org/abs/astro-ph/0607224}
  {arXiv:astro-ph/0607224 [astro-ph]} \BibitemShut {NoStop}%
%%CITATION = ASTRO-PH/0607224;%%
\bibitem [{\citenamefont {Zel'Dovich}\ and\ \citenamefont
  {Polnarev}(1974)}]{zel1974radiation}%
  \BibitemOpen
  \bibfield  {author} {\bibinfo {author} {\bibfnamefont {Y.~B.}\ \bibnamefont
  {Zel'Dovich}}\ and\ \bibinfo {author} {\bibfnamefont {A.}~\bibnamefont
  {Polnarev}},\ }\href@noop {} {\bibfield  {journal} {\bibinfo  {journal}
  {Soviet Astr}\ }\textbf {\bibinfo {volume} {18}},\ \bibinfo {pages} {17}
  (\bibinfo {year} {1974})}\BibitemShut {NoStop}%
\bibitem [{\citenamefont {Smarr}(1977)}]{Smarr:1977fy}%
  \BibitemOpen
  \bibfield  {author} {\bibinfo {author} {\bibfnamefont {L.}~\bibnamefont
  {Smarr}},\ }\href {\doibase 10.1103/PhysRevD.15.2069} {\bibfield  {journal}
  {\bibinfo  {journal} {Phys. Rev.}\ }\textbf {\bibinfo {volume} {D15}},\
  \bibinfo {pages} {2069} (\bibinfo {year} {1977})}\BibitemShut {NoStop}%
%%CITATION = PHRVA,D15,2069;%%
\bibitem [{\citenamefont {Kovacs}\ and\ \citenamefont
  {Thorne}(1978)}]{Kovacs:1978eu}%
  \BibitemOpen
  \bibfield  {author} {\bibinfo {author} {\bibfnamefont {S.~J.}\ \bibnamefont
  {Kovacs}}\ and\ \bibinfo {author} {\bibfnamefont {K.~S.}\ \bibnamefont
  {Thorne}},\ }\href {\doibase 10.1086/156350} {\bibfield  {journal} {\bibinfo
  {journal} {Astrophys. J.}\ }\textbf {\bibinfo {volume} {224}},\ \bibinfo
  {pages} {62} (\bibinfo {year} {1978})}\BibitemShut {NoStop}%
%%CITATION = ASJOA,224,62;%%
\bibitem [{\citenamefont {Braginskii}\ and\ \citenamefont
  {Grishchuk}(1985)}]{Braginskii:1985}%
  \BibitemOpen
  \bibfield  {author} {\bibinfo {author} {\bibfnamefont {V.~B.}\ \bibnamefont
  {Braginskii}}\ and\ \bibinfo {author} {\bibfnamefont {L.}~\bibnamefont
  {Grishchuk}},\ }\href@noop {} {\bibfield  {journal} {\bibinfo  {journal}
  {Sov. Phys. JETP}\ }\textbf {\bibinfo {volume} {62}},\ \bibinfo {pages} {427}
  (\bibinfo {year} {1985})}\BibitemShut {NoStop}%
\bibitem [{\citenamefont {Turner}(1977)}]{Turner:1977}%
  \BibitemOpen
  \bibfield  {author} {\bibinfo {author} {\bibfnamefont {M.~S.}\ \bibnamefont
  {Turner}},\ }\href {\doibase 10.1086/155501} {\bibfield  {journal} {\bibinfo
  {journal} {Astrophys. J.}\ }\textbf {\bibinfo {volume} {216}},\ \bibinfo
  {pages} {610} (\bibinfo {year} {1977})}\BibitemShut {NoStop}%
\bibitem [{\citenamefont {Epstein}(1978)}]{Epstein:1978dv}%
  \BibitemOpen
  \bibfield  {author} {\bibinfo {author} {\bibfnamefont {R.}~\bibnamefont
  {Epstein}},\ }\href {\doibase 10.1086/156337} {\bibfield  {journal} {\bibinfo
   {journal} {Astrophys. J.}\ }\textbf {\bibinfo {volume} {223}},\ \bibinfo
  {pages} {1037} (\bibinfo {year} {1978})}\BibitemShut {NoStop}%
%%CITATION = ASJOA,223,1037;%%
\bibitem [{\citenamefont {Turner}(1978)}]{Turner:1978jj}%
  \BibitemOpen
  \bibfield  {author} {\bibinfo {author} {\bibfnamefont {M.~S.}\ \bibnamefont
  {Turner}},\ }\href {\doibase 10.1038/274565a0} {\bibfield  {journal}
  {\bibinfo  {journal} {Nature}\ }\textbf {\bibinfo {volume} {274}},\ \bibinfo
  {pages} {565} (\bibinfo {year} {1978})}\BibitemShut {NoStop}%
%%CITATION = NATUA,274,565;%%
\bibitem [{\citenamefont {Braginskii}\ and\ \citenamefont
  {Thorne}(1987)}]{Braginskii:1987}%
  \BibitemOpen
  \bibfield  {author} {\bibinfo {author} {\bibfnamefont {V.~B.}\ \bibnamefont
  {Braginskii}}\ and\ \bibinfo {author} {\bibfnamefont {K.~S.}\ \bibnamefont
  {Thorne}},\ }\href {\doibase 10.1038/327123a0} {\bibfield  {journal}
  {\bibinfo  {journal} {Nature}\ }\textbf {\bibinfo {volume} {327}},\ \bibinfo
  {pages} {23} (\bibinfo {year} {1987})}\BibitemShut {NoStop}%
\bibitem [{\citenamefont {Favata}(2010)}]{Favata:2010zu}%
  \BibitemOpen
  \bibfield  {author} {\bibinfo {author} {\bibfnamefont {M.}~\bibnamefont
  {Favata}},\ }\href {\doibase 10.1088/0264-9381/27/8/084036} {\bibfield
  {journal} {\bibinfo  {journal} {Class. Quant. Grav.}\ }\textbf {\bibinfo
  {volume} {27}},\ \bibinfo {pages} {084036} (\bibinfo {year} {2010})},\
  \Eprint {http://arxiv.org/abs/1003.3486} {arXiv:1003.3486 [gr-qc]}
  \BibitemShut {NoStop}%
%%CITATION = ARXIV:1003.3486;%%
\bibitem [{\citenamefont {Sago}\ \emph {et~al.}(2004)\citenamefont {Sago},
  \citenamefont {Ioka}, \citenamefont {Nakamura},\ and\ \citenamefont
  {Yamazaki}}]{Sago:2004pn}%
  \BibitemOpen
  \bibfield  {author} {\bibinfo {author} {\bibfnamefont {N.}~\bibnamefont
  {Sago}}, \bibinfo {author} {\bibfnamefont {K.}~\bibnamefont {Ioka}}, \bibinfo
  {author} {\bibfnamefont {T.}~\bibnamefont {Nakamura}}, \ and\ \bibinfo
  {author} {\bibfnamefont {R.}~\bibnamefont {Yamazaki}},\ }\href {\doibase
  10.1103/PhysRevD.70.104012} {\bibfield  {journal} {\bibinfo  {journal} {Phys.
  Rev.}\ }\textbf {\bibinfo {volume} {D70}},\ \bibinfo {pages} {104012}
  (\bibinfo {year} {2004})},\ \Eprint {http://arxiv.org/abs/gr-qc/0405067}
  {arXiv:gr-qc/0405067 [gr-qc]} \BibitemShut {NoStop}%
%%CITATION = GR-QC/0405067;%%
\bibitem [{\citenamefont {Thorne}(1987)}]{300yearsofgravitation}%
  \BibitemOpen
  \bibfield  {author} {\bibinfo {author} {\bibfnamefont {K.}~\bibnamefont
  {Thorne}},\ }in\ \href@noop {} {\emph {\bibinfo {booktitle} {Three hundred
  years of gravitation~\text{\normalfont, edited by S. W. Hawking and W.
  Isarel}}}}\ (\bibinfo  {publisher} {Cambridge University Press},\ \bibinfo
  {year} {1987})\ pp.\ \bibinfo {pages} {330--458}\BibitemShut {NoStop}%
\bibitem [{\citenamefont {Shi}\ and\ \citenamefont
  {Fuller}(1998)}]{Shi:1998nd}%
  \BibitemOpen
  \bibfield  {author} {\bibinfo {author} {\bibfnamefont {X.-D.}\ \bibnamefont
  {Shi}}\ and\ \bibinfo {author} {\bibfnamefont {G.~M.}\ \bibnamefont
  {Fuller}},\ }\href {\doibase 10.1086/305992} {\bibfield  {journal} {\bibinfo
  {journal} {Astrophys. J.}\ }\textbf {\bibinfo {volume} {503}},\ \bibinfo
  {pages} {307} (\bibinfo {year} {1998})},\ \Eprint
  {http://arxiv.org/abs/astro-ph/9801106} {arXiv:astro-ph/9801106 [astro-ph]}
  \BibitemShut {NoStop}%
%%CITATION = ASTRO-PH/9801106;%%
\bibitem [{\citenamefont {Schinder}\ \emph {et~al.}(1987)\citenamefont
  {Schinder}, \citenamefont {Schramm}, \citenamefont {Wiita}, \citenamefont
  {Margolis},\ and\ \citenamefont {Tubbs}}]{Schinder:1986nh}%
  \BibitemOpen
  \bibfield  {author} {\bibinfo {author} {\bibfnamefont {P.~J.}\ \bibnamefont
  {Schinder}}, \bibinfo {author} {\bibfnamefont {D.~N.}\ \bibnamefont
  {Schramm}}, \bibinfo {author} {\bibfnamefont {P.~J.}\ \bibnamefont {Wiita}},
  \bibinfo {author} {\bibfnamefont {S.~H.}\ \bibnamefont {Margolis}}, \ and\
  \bibinfo {author} {\bibfnamefont {D.~L.}\ \bibnamefont {Tubbs}},\ }\href
  {\doibase 10.1086/164993} {\bibfield  {journal} {\bibinfo  {journal}
  {Astrophys. J.}\ }\textbf {\bibinfo {volume} {313}},\ \bibinfo {pages} {531}
  (\bibinfo {year} {1987})}\BibitemShut {NoStop}%
%%CITATION = ASJOA,313,531;%%
\bibitem [{\citenamefont {Itoh}\ \emph {et~al.}(1989)\citenamefont {Itoh},
  \citenamefont {Adachi}, \citenamefont {Nakagawa}, \citenamefont {Kohyama},\
  and\ \citenamefont {Munakata}}]{Itoh1989}%
  \BibitemOpen
  \bibfield  {author} {\bibinfo {author} {\bibfnamefont {N.}~\bibnamefont
  {Itoh}}, \bibinfo {author} {\bibfnamefont {T.}~\bibnamefont {Adachi}},
  \bibinfo {author} {\bibfnamefont {M.}~\bibnamefont {Nakagawa}}, \bibinfo
  {author} {\bibfnamefont {Y.}~\bibnamefont {Kohyama}}, \ and\ \bibinfo
  {author} {\bibfnamefont {H.}~\bibnamefont {Munakata}},\ }\href {\doibase
  10.1086/167301} {\bibfield  {journal} {\bibinfo  {journal} {Astrophys. J.}\
  }\textbf {\bibinfo {volume} {339}},\ \bibinfo {pages} {354} (\bibinfo {year}
  {1989})}\BibitemShut {NoStop}%
\bibitem [{\citenamefont {Itoh}\ \emph {et~al.}(1996)\citenamefont {Itoh},
  \citenamefont {Hayashi}, \citenamefont {Nishikawa},\ and\ \citenamefont
  {Kohyama}}]{Itoh1996}%
  \BibitemOpen
  \bibfield  {author} {\bibinfo {author} {\bibfnamefont {N.}~\bibnamefont
  {Itoh}}, \bibinfo {author} {\bibfnamefont {H.}~\bibnamefont {Hayashi}},
  \bibinfo {author} {\bibfnamefont {A.}~\bibnamefont {Nishikawa}}, \ and\
  \bibinfo {author} {\bibfnamefont {Y.}~\bibnamefont {Kohyama}},\ }\href
  {\doibase 10.1086/192264} {\bibfield  {journal} {\bibinfo  {journal}
  {Astrophys. J. S.}\ }\textbf {\bibinfo {volume} {102}},\ \bibinfo {pages}
  {411} (\bibinfo {year} {1996})}\BibitemShut {NoStop}%
\bibitem [{\citenamefont {Linke}\ \emph {et~al.}(2001)\citenamefont {Linke},
  \citenamefont {Font}, \citenamefont {Janka}, \citenamefont {Muller},\ and\
  \citenamefont {Papadopoulos}}]{Linke:2001mq}%
  \BibitemOpen
  \bibfield  {author} {\bibinfo {author} {\bibfnamefont {F.}~\bibnamefont
  {Linke}}, \bibinfo {author} {\bibfnamefont {J.~A.}\ \bibnamefont {Font}},
  \bibinfo {author} {\bibfnamefont {H.-T.}\ \bibnamefont {Janka}}, \bibinfo
  {author} {\bibfnamefont {E.}~\bibnamefont {Muller}}, \ and\ \bibinfo {author}
  {\bibfnamefont {P.}~\bibnamefont {Papadopoulos}},\ }\href {\doibase
  10.1051/0004-6361:20010993} {\bibfield  {journal} {\bibinfo  {journal}
  {Astron. Astrophys.}\ }\textbf {\bibinfo {volume} {376}},\ \bibinfo {pages}
  {568} (\bibinfo {year} {2001})},\ \Eprint
  {http://arxiv.org/abs/astro-ph/0103144} {arXiv:astro-ph/0103144 [astro-ph]}
  \BibitemShut {NoStop}%
%%CITATION = ASTRO-PH/0103144;%%
\bibitem [{\citenamefont {Chen}\ \emph {et~al.}(2014)\citenamefont {Chen},
  \citenamefont {Heger}, \citenamefont {Woosley}, \citenamefont {Almgren},
  \citenamefont {Whalen},\ and\ \citenamefont {Johnson}}]{Chen:2014yea}%
  \BibitemOpen
  \bibfield  {author} {\bibinfo {author} {\bibfnamefont {K.-J.}\ \bibnamefont
  {Chen}}, \bibinfo {author} {\bibfnamefont {A.}~\bibnamefont {Heger}},
  \bibinfo {author} {\bibfnamefont {S.}~\bibnamefont {Woosley}}, \bibinfo
  {author} {\bibfnamefont {A.}~\bibnamefont {Almgren}}, \bibinfo {author}
  {\bibfnamefont {D.}~\bibnamefont {Whalen}}, \ and\ \bibinfo {author}
  {\bibfnamefont {J.}~\bibnamefont {Johnson}},\ }\href {\doibase
  10.1088/0004-637X/790/2/162} {\bibfield  {journal} {\bibinfo  {journal}
  {Astrophys. J.}\ }\textbf {\bibinfo {volume} {790}},\ \bibinfo {pages} {162}
  (\bibinfo {year} {2014})},\ \Eprint {http://arxiv.org/abs/1402.4777}
  {arXiv:1402.4777 [astro-ph.HE]} \BibitemShut {NoStop}%
%%CITATION = ARXIV:1402.4777;%%
\bibitem [{\citenamefont {Sachs}\ and\ \citenamefont
  {Wolfe}(1967)}]{Sachs:1967er}%
  \BibitemOpen
  \bibfield  {author} {\bibinfo {author} {\bibfnamefont {R.~K.}\ \bibnamefont
  {Sachs}}\ and\ \bibinfo {author} {\bibfnamefont {A.~M.}\ \bibnamefont
  {Wolfe}},\ }\href {\doibase 10.1007/s10714-007-0448-9} {\bibfield  {journal}
  {\bibinfo  {journal} {Astrophys. J.}\ }\textbf {\bibinfo {volume} {147}},\
  \bibinfo {pages} {73} (\bibinfo {year} {1967})},\ \bibinfo {note} {[Gen. Rel.
  Grav.39,1929(2007)]}\BibitemShut {NoStop}%
%%CITATION = ASJOA,147,73;%%
\bibitem [{\citenamefont {Cardall}\ and\ \citenamefont
  {Fuller}(1997)}]{Cardall:1997bi}%
  \BibitemOpen
  \bibfield  {author} {\bibinfo {author} {\bibfnamefont {C.~Y.}\ \bibnamefont
  {Cardall}}\ and\ \bibinfo {author} {\bibfnamefont {G.~M.}\ \bibnamefont
  {Fuller}},\ }\href {\doibase 10.1086/310838} {\bibfield  {journal} {\bibinfo
  {journal} {Astrophys. J.}\ }\textbf {\bibinfo {volume} {486}},\ \bibinfo
  {pages} {L111} (\bibinfo {year} {1997})},\ \Eprint
  {http://arxiv.org/abs/astro-ph/9701178} {arXiv:astro-ph/9701178 [astro-ph]}
  \BibitemShut {NoStop}%
%%CITATION = ASTRO-PH/9701178;%%
\bibitem [{\citenamefont {Apostolatos}\ \emph {et~al.}(1994)\citenamefont
  {Apostolatos}, \citenamefont {Cutler}, \citenamefont {Sussman},\ and\
  \citenamefont {Thorne}}]{Apostolatos:1994mx}%
  \BibitemOpen
  \bibfield  {author} {\bibinfo {author} {\bibfnamefont {T.~A.}\ \bibnamefont
  {Apostolatos}}, \bibinfo {author} {\bibfnamefont {C.}~\bibnamefont {Cutler}},
  \bibinfo {author} {\bibfnamefont {G.~J.}\ \bibnamefont {Sussman}}, \ and\
  \bibinfo {author} {\bibfnamefont {K.~S.}\ \bibnamefont {Thorne}},\ }\href
  {\doibase 10.1103/PhysRevD.49.6274} {\bibfield  {journal} {\bibinfo
  {journal} {Phys. Rev.}\ }\textbf {\bibinfo {volume} {D49}},\ \bibinfo {pages}
  {6274} (\bibinfo {year} {1994})}\BibitemShut {NoStop}%
%%CITATION = PHRVA,D49,6274;%%
\bibitem [{\citenamefont {Barack}\ and\ \citenamefont
  {Cutler}(2004)}]{Barack:2003fp}%
  \BibitemOpen
  \bibfield  {author} {\bibinfo {author} {\bibfnamefont {L.}~\bibnamefont
  {Barack}}\ and\ \bibinfo {author} {\bibfnamefont {C.}~\bibnamefont
  {Cutler}},\ }\href {\doibase 10.1103/PhysRevD.69.082005} {\bibfield
  {journal} {\bibinfo  {journal} {Phys. Rev.}\ }\textbf {\bibinfo {volume}
  {D69}},\ \bibinfo {pages} {082005} (\bibinfo {year} {2004})},\ \Eprint
  {http://arxiv.org/abs/gr-qc/0310125} {arXiv:gr-qc/0310125 [gr-qc]}
  \BibitemShut {NoStop}%
%%CITATION = GR-QC/0310125;%%
\bibitem [{\citenamefont {Yagi}\ and\ \citenamefont
  {Seto}(2011)}]{Yagi:2011wg}%
  \BibitemOpen
  \bibfield  {author} {\bibinfo {author} {\bibfnamefont {K.}~\bibnamefont
  {Yagi}}\ and\ \bibinfo {author} {\bibfnamefont {N.}~\bibnamefont {Seto}},\
  }\href {\doibase 10.1103/PhysRevD.95.109901, 10.1103/PhysRevD.83.044011}
  {\bibfield  {journal} {\bibinfo  {journal} {Phys. Rev.}\ }\textbf {\bibinfo
  {volume} {D83}},\ \bibinfo {pages} {044011} (\bibinfo {year} {2011})},\
  \bibinfo {note} {[Erratum: Phys. Rev.D95,no.10,109901(2017)]},\ \Eprint
  {http://arxiv.org/abs/1101.3940} {arXiv:1101.3940 [astro-ph.CO]} \BibitemShut
  {NoStop}%
%%CITATION = ARXIV:1101.3940;%%
\bibitem [{\citenamefont {Payne}(1983)}]{Payne:1984ec}%
  \BibitemOpen
  \bibfield  {author} {\bibinfo {author} {\bibfnamefont {P.~N.}\ \bibnamefont
  {Payne}},\ }\href {\doibase 10.1103/PhysRevD.28.1894} {\bibfield  {journal}
  {\bibinfo  {journal} {Phys. Rev.}\ }\textbf {\bibinfo {volume} {D28}},\
  \bibinfo {pages} {1894} (\bibinfo {year} {1983})}\BibitemShut {NoStop}%
%%CITATION = PHRVA,D28,1894;%%
\bibitem [{\citenamefont {Foster}\ and\ \citenamefont
  {Backer}(1990)}]{foster1990constructing}%
  \BibitemOpen
  \bibfield  {author} {\bibinfo {author} {\bibfnamefont {R.~S.}\ \bibnamefont
  {Foster}}\ and\ \bibinfo {author} {\bibfnamefont {D.~C.}\ \bibnamefont
  {Backer}},\ }\href@noop {} {\bibfield  {journal} {\bibinfo  {journal}
  {Astrophys. J.}\ }\textbf {\bibinfo {volume} {361}},\ \bibinfo {pages} {300}
  (\bibinfo {year} {1990})}\BibitemShut {NoStop}%
\bibitem [{\citenamefont {Lorimer}(2008)}]{lorimer2008binary}%
  \BibitemOpen
  \bibfield  {author} {\bibinfo {author} {\bibfnamefont {D.~R.}\ \bibnamefont
  {Lorimer}},\ }\href@noop {} {\bibfield  {journal} {\bibinfo  {journal}
  {Living reviews in relativity}\ }\textbf {\bibinfo {volume} {11}},\ \bibinfo
  {pages} {8} (\bibinfo {year} {2008})}\BibitemShut {NoStop}%
\bibitem [{\citenamefont {Jenet}\ \emph {et~al.}(2011)\citenamefont {Jenet},
  \citenamefont {Armstrong},\ and\ \citenamefont {Tinto}}]{Jenet:2011me}%
  \BibitemOpen
  \bibfield  {author} {\bibinfo {author} {\bibfnamefont {F.~A.}\ \bibnamefont
  {Jenet}}, \bibinfo {author} {\bibfnamefont {J.~W.}\ \bibnamefont
  {Armstrong}}, \ and\ \bibinfo {author} {\bibfnamefont {M.}~\bibnamefont
  {Tinto}},\ }\href {\doibase 10.1103/PhysRevD.83.081301} {\bibfield  {journal}
  {\bibinfo  {journal} {Phys. Rev.}\ }\textbf {\bibinfo {volume} {D83}},\
  \bibinfo {pages} {081301} (\bibinfo {year} {2011})},\ \Eprint
  {http://arxiv.org/abs/1101.3759} {arXiv:1101.3759 [gr-qc]} \BibitemShut
  {NoStop}%
%%CITATION = ARXIV:1101.3759;%%
\bibitem [{\citenamefont {Harry}\ \emph {et~al.}(2006)\citenamefont {Harry},
  \citenamefont {Fritschel}, \citenamefont {Shaddock}, \citenamefont
  {Folkner},\ and\ \citenamefont {Phinney}}]{Harry:2006fi}%
  \BibitemOpen
  \bibfield  {author} {\bibinfo {author} {\bibfnamefont {G.~M.}\ \bibnamefont
  {Harry}}, \bibinfo {author} {\bibfnamefont {P.}~\bibnamefont {Fritschel}},
  \bibinfo {author} {\bibfnamefont {D.~A.}\ \bibnamefont {Shaddock}}, \bibinfo
  {author} {\bibfnamefont {W.}~\bibnamefont {Folkner}}, \ and\ \bibinfo
  {author} {\bibfnamefont {E.~S.}\ \bibnamefont {Phinney}},\ }\href {\doibase
  10.1088/0264-9381/23/24/C01, 10.1088/0264-9381/23/15/008} {\bibfield
  {journal} {\bibinfo  {journal} {Class. Quant. Grav.}\ }\textbf {\bibinfo
  {volume} {23}},\ \bibinfo {pages} {4887} (\bibinfo {year} {2006})},\ \bibinfo
  {note} {[Erratum: Class. Quant. Grav.23,7361(2006)]}\BibitemShut {NoStop}%
%%CITATION = CQGRD,23,4887;%%
\bibitem [{\citenamefont {Crowder}\ and\ \citenamefont
  {Cornish}(2005)}]{Crowder:2005nr}%
  \BibitemOpen
  \bibfield  {author} {\bibinfo {author} {\bibfnamefont {J.}~\bibnamefont
  {Crowder}}\ and\ \bibinfo {author} {\bibfnamefont {N.~J.}\ \bibnamefont
  {Cornish}},\ }\href {\doibase 10.1103/PhysRevD.72.083005} {\bibfield
  {journal} {\bibinfo  {journal} {Phys. Rev.}\ }\textbf {\bibinfo {volume}
  {D72}},\ \bibinfo {pages} {083005} (\bibinfo {year} {2005})},\ \Eprint
  {http://arxiv.org/abs/gr-qc/0506015} {arXiv:gr-qc/0506015 [gr-qc]}
  \BibitemShut {NoStop}%
%%CITATION = GR-QC/0506015;%%
\bibitem [{\citenamefont {Audley}\ \emph {et~al.}(2017)\citenamefont {Audley}
  \emph {et~al.}}]{Audley:2017drz}%
  \BibitemOpen
  \bibfield  {author} {\bibinfo {author} {\bibfnamefont {H.}~\bibnamefont
  {Audley}} \emph {et~al.},\ }\href@noop {} {\  (\bibinfo {year} {2017})},\
  \Eprint {http://arxiv.org/abs/1702.00786} {arXiv:1702.00786 [astro-ph.IM]}
  \BibitemShut {NoStop}%
%%CITATION = ARXIV:1702.00786;%%
\bibitem [{\citenamefont {Montero}\ \emph {et~al.}(2012)\citenamefont
  {Montero}, \citenamefont {Janka},\ and\ \citenamefont
  {Muller}}]{Montero:2011ps}%
  \BibitemOpen
  \bibfield  {author} {\bibinfo {author} {\bibfnamefont {P.~J.}\ \bibnamefont
  {Montero}}, \bibinfo {author} {\bibfnamefont {H.-T.}\ \bibnamefont {Janka}},
  \ and\ \bibinfo {author} {\bibfnamefont {E.}~\bibnamefont {Muller}},\ }\href
  {\doibase 10.1088/0004-637X/749/1/37} {\bibfield  {journal} {\bibinfo
  {journal} {Astrophys. J.}\ }\textbf {\bibinfo {volume} {749}},\ \bibinfo
  {pages} {37} (\bibinfo {year} {2012})},\ \Eprint
  {http://arxiv.org/abs/1108.3090} {arXiv:1108.3090 [astro-ph.CO]} \BibitemShut
  {NoStop}%
%%CITATION = ARXIV:1108.3090;%%
\bibitem [{\citenamefont {Fuller}\ and\ \citenamefont
  {Shi}(1998)}]{Fuller:1997em}%
  \BibitemOpen
  \bibfield  {author} {\bibinfo {author} {\bibfnamefont {G.~M.}\ \bibnamefont
  {Fuller}}\ and\ \bibinfo {author} {\bibfnamefont {X.-D.}\ \bibnamefont
  {Shi}},\ }\href {\doibase 10.1086/311477} {\bibfield  {journal} {\bibinfo
  {journal} {Astrophys. J.}\ }\textbf {\bibinfo {volume} {502}},\ \bibinfo
  {pages} {L5} (\bibinfo {year} {1998})},\ \Eprint
  {http://arxiv.org/abs/astro-ph/9711020} {arXiv:astro-ph/9711020 [astro-ph]}
  \BibitemShut {NoStop}%
%%CITATION = ASTRO-PH/9711020;%%
\bibitem [{\citenamefont {Shi}\ \emph {et~al.}(1998)\citenamefont {Shi},
  \citenamefont {Fuller},\ and\ \citenamefont {Halzen}}]{Shi:1998jx}%
  \BibitemOpen
  \bibfield  {author} {\bibinfo {author} {\bibfnamefont {X.-D.}\ \bibnamefont
  {Shi}}, \bibinfo {author} {\bibfnamefont {G.~M.}\ \bibnamefont {Fuller}}, \
  and\ \bibinfo {author} {\bibfnamefont {F.}~\bibnamefont {Halzen}},\ }\href
  {\doibase 10.1103/PhysRevLett.81.5722} {\bibfield  {journal} {\bibinfo
  {journal} {Phys. Rev. Lett.}\ }\textbf {\bibinfo {volume} {81}},\ \bibinfo
  {pages} {5722} (\bibinfo {year} {1998})},\ \Eprint
  {http://arxiv.org/abs/astro-ph/9805242} {arXiv:astro-ph/9805242 [astro-ph]}
  \BibitemShut {NoStop}%
%%CITATION = ASTRO-PH/9805242;%%
\end{thebibliography}%

\end{document}